\documentclass[twocolumn,secnumarabic,nobibnotes,aps,nofootinbib]{revtex4-1}

\usepackage{color, xcolor, colortbl}
\usepackage{graphicx,epstopdf}
\usepackage{amsmath,amssymb,amsthm}
\usepackage{bm}
\usepackage{appendix}
\usepackage{multirow}
\usepackage{braket}
\usepackage[english]{babel}
\usepackage{hyperref}
\usepackage[capitalize]{cleveref}
\usepackage{xspace}
\usepackage[qm]{qcircuit}
\usepackage{subfigure}
\usepackage{multirow}

\usepackage{newfloat,algcompatible}
\usepackage[size=small,justification=centerlast]{caption}
\usepackage{etoolbox}

\AtEndEnvironment{algorithm}{\noindent\hrulefill\par\nobreak\vskip-8pt}

\DeclareFloatingEnvironment[
fileext=loa,
listname=List of Algorithms,
name=ALGORITHM,
placement=tbhp,
]{algorithm}
\DeclareCaptionFormat{algorithms}{\vskip-6pt\hrulefill\par#1#2#3\vskip-6pt\hrulefill}
\algblock[Input]{Input}{EndInput}
\algblockdefx[Input]{Input}{EndInput}
[1]{\textbf{Input} #1}
{}

\renewcommand{\Re}{\operatorname{Re}}
\renewcommand{\Im}{\operatorname{Im}}

\newcommand{\Tr}{\operatorname{Tr}}

\newcommand{\I}{\mathrm{i}}

\newcommand{\mc}[1]{\mathcal{#1}}

\newcommand{\wt}[1]{\widetilde{#1}}

\newcommand{\abs}[1]{\left\lvert#1\right\rvert}
\newcommand{\norm}[1]{\left\lVert#1\right\rVert}

\newcommand{\Or}{\mathcal{O}}

\newcommand{\CC}{\mathbb{C}}

\theoremstyle{plain}

\theoremstyle{remark}

\theoremstyle{plain}
\newtheorem*{lem*}{\protect\lemmaname}
\theoremstyle{plain}

\theoremstyle{plain}

\providecommand{\corollaryname}{Corollary}
\providecommand{\lemmaname}{Lemma}
\providecommand{\propositionname}{Proposition}
\providecommand{\remarkname}{Remark}
\providecommand{\theoremname}{Theorem}

\newcommand{\bem}{{RACBEM}\xspace}
\newcommand{\hbem}{{H-RACBEM}\xspace}
\newcommand{\chbem}{{canonical H-RACBEM}\xspace}
\newcommand{\qspdeg}{d}
\newcommand{\scalefac}{\alpha}

\newcommand{\BQIC}{Berkeley Center for Quantum Information and Computation, Berkeley, California 94720 USA}
\newcommand{\DeptChem}{Department of Chemistry, University of California, Berkeley, California 94720 USA}
\newcommand{\DeptMath}{Department of Mathematics, University of California, Berkeley, California 94720 USA}
\newcommand{\LBLMath}{Computational Research Division, Lawrence Berkeley National Laboratory, Berkeley, CA 94720, USA}

\begin{document}

\title{Random circuit block-encoded matrix and a proposal of quantum LINPACK benchmark}

\author{Yulong Dong$^{1,2}$} 
\author{Lin Lin$^{3,4}$}
\email[Electronic address: ]{linlin@math.berkeley.edu}

\affiliation{$^1$\BQIC}
\affiliation{$^2$\DeptChem}
\affiliation{$^3$\DeptMath}
\affiliation{$^4$\LBLMath}

\begin{abstract}
	
The LINPACK benchmark reports the performance of a computer for solving a system of linear equations with dense random matrices. Although this task was not designed with a real application directly in mind, the LINPACK benchmark has been used to define the list of TOP500 supercomputers since the debut of the list in 1993. We propose that a similar benchmark, called the quantum LINPACK benchmark, could be used to measure the whole machine performance of quantum computers. The success of the quantum LINPACK benchmark should be viewed as the minimal requirement for a quantum computer to perform a useful task of solving linear algebra problems, such as linear systems of equations. We propose an input model called the RAndom Circuit Block-Encoded Matrix (\bem), which is a proper generalization of a dense random matrix in the quantum setting. The RACBEM model is efficient to be implemented on a quantum computer, and can be designed to optimally adapt to any given quantum architecture, with relying on a black-box quantum compiler. Besides solving linear systems, the RACBEM model can be used to perform a variety of linear algebra tasks relevant to many physical applications, such as computing spectral measures, time series generated by a Hamiltonian simulation, and thermal averages of the energy. We implement these linear algebra operations on IBM Q quantum devices as well as quantum virtual machines, and demonstrate their performance in solving scientific computing problems.

\end{abstract}

\maketitle

\section{Introduction}

Quantum computers hold the promise of dramatically accelerating calculations in a wide range of fields, and quantum supremacy was achieved in 2019 via sampling random quantum circuits \cite{AruteAryaBabbushEtAl2019}. Assume that
there are ten thousand quantum computers (or many more) available now, how should we select the top 500 best performing computers for scientific computing applications? The answer in the context of classical supercomputers is given by the LINPACK benchmark \cite{DongarraLuszczekPetitet2003}, which measures the floating point computing power of a classical computer via its performance for solving linear systems of equations $Ax=b$. The input matrix $A$ is a dense pseudo-random matrix, and there is no immediate application associated with such a matrix (similar to a quantum supremacy experiment in this sense). There has been much controversy over its effectiveness in measuring the capability of classical computers in scientific computing applications since the very beginning. However, LINPACK is widely used and performance numbers are available for almost all relevant systems.
The LINPACK benchmark has been used as the defining criterion of TOP500 supercomputers since the debut of the list in 1993 \cite{TOP500}.
One important reason is that dense matrices and dense matrix operations are relatively easy to implement and to optimize on classical computers. These operations have been tuned to be highly scalable, which enabled the performance benchmark of systems that cover a performance range of 12 orders of magnitude in the past 20 years. 

In order to mimic the success of the LINPACK benchmark on quantum computers, we consider the problem of solving the quantum linear system problem (QLSP). Many challenging high-dimensional problems in physics, such as computing Green's functions for a quantum many-body system, can be formulated in terms of QLSP. This field has witnessed significant progresses in the past few years \cite{HarrowHassidimLloyd2009,CaoPapageorgiouPetrasEtAl2013,ChildsKothariSomma2017,ChakrabortyGilyenJeffery2018,GilyenSuLowEtAl2019,SubasiSommaOrsucci2019,WossnigZhaoPrakash2018,AnLin2019,LinTong2019,XuSunEndoEtAl2019,Bravo-PrietoLaRoseCerezoEtAl2019,CasaresMartin-Delgado2019,TongAnWiebeEtAl2020}. Shortly speaking, given $A\in \CC^{2^n\times 2^n}$ and $\ket{b}\in\CC^{2^n}$, QLSP aims at obtaining an $n$-qubit solution vector $\ket{x}\propto A^{-1}\ket{b}$.
More precisely and using the language of block-encoding \cite{GilyenSuLowEtAl2019}, QLSP is the problem of finding an $(m+n)$-qubit unitary matrix $U$, such that 
\begin{equation}
\ket{x}=(\bra{0^m}\otimes I_n)U(\ket{0^m}\otimes\ket{b})=\frac{A^{-1}\ket{b}}{\norm{A^{-1}\ket{b}}}.
\label{eqn:QLSP}
\end{equation}
In other words, the solution is obtained upon measuring $0$ for all  $m$ ancilla qubits (with a success probability $p)$.

In this paper, we propose the quantum LINPACK benchmark, which can be concisely stated as the problem of \emph{using the quantum computer to evaluate the success probability $p$ for a certain random matrix $A$}. While the rationale of such a task will be discussed in detail in the main text, we first emphasize that the quantum LINPACK benchmark examines the \textit{whole machine performance} of quantum computers, rather than the performance of a few qubits as often measured by methods such as randomized benchmarking \cite{MagesanGambettaEmerson2011} and gateset tomography \cite{Blume-KohoutGambleNielsenEtAl2017}. There have been a number of whole machine quantum benchmarks proposed in the literature, such as quantum volume \cite{CrossBishopSheldonEtAl2019}, cycle benchmarking \cite{ErhardWallmanPostlerEtAl2019}, and the linear cross-entropy benchmarking as in the supremacy experiment \cite{BoixoIsakovSmelyanskiyEtAl2018,AruteAryaBabbushEtAl2019}. However, those benchmark methods are proposed for generic settings and therefore they are less representative for the performance on the user-specified applications. In particular, the error of a structured circuit can be significantly larger than that of a fully randomized one \cite{ProctorRudingerYoungEtAl2020}. 
The quantum LINPACK benchmark targets directly at the performance of quantum computers for scientific computing applications, as in the case of the LINPACK benchmark for classical supercomputers. We emphasize that the success of the quantum LINPACK benchmark does not guarantee that the quantum computer has solved $\ket{x}$ accurately, but should be viewed as the minimal requirement for a quantum computer to solve QLSP. Given the wide range of potential applications of QLSP from quantum many body problems to quantum machine learning, it is important for future quantum computers to first meet the criterion of the quantum LINPACK benchmark, in order to achieve quantum advantage via the path of solving linear algebra problems.

In order to perform the quantum LINPACK benchmark, note that it would be highly inefficient if we first generate a dense pseudo-random matrix $A$ classically and then feed it into the quantum computer using e.g. QRAM \cite{GiovannettiLloydMaccone2008}. In fact, such a strong assumption on the input model often lead to dequantized classical algorithms \cite{Tang2019}. Instead we focus on matrices that are \emph{inherently easy} to generate on quantum computers. In particular, the supremacy experiment inspires us to generate a random matrix directly using a random quantum circuit. 

We propose an input model called the RAndom Circuit Block-Encoded Matrix (\bem). We argue that the \bem model is a proper generalization of dense random matrices in the quantum setting, suitable for linear algebra tasks. The \bem model, and its Hermitian version called \hbem model, are simple to construct and allow us to get access to in principle any $n$-qubit matrix and $n$-qubit Hermitian matrix, respectively (up to a scaling factor) by adding only one ancilla qubit. Together with the recently developed technique of quantum singular value transformation (QSVT) \cite{GilyenSuLowEtAl2019}, we yield a practical algorithm for performing the quantum LINPACK benchmark on near-term devices with a shallow circuit depth.

With QSVT and the \hbem model, the circuit used in the quantum LINPACK benchmark can be designed to adapt to the coupling map of almost any given gate-based quantum architecture. All operations can be carried out with straightforward usage of basic one-qubit gates and CNOT gates, and there is no complex controlled unitaries involved. Due to the use of the basic gate set and the adaptivity to the quantum architecture, the quantum LINPACK benchmark does not require the explicit use of the compiler, while the randomized benchmarking requires gate compilation. Furthermore, the number of ancilla qubits needed is minimal (usually 2). By using the \hbem model, the condition number of the random matrices is fully controllable which is crucial for reducing the circuit depth. While the development of quantum hardware has been very rapid in the past few years, quantum resources are expected to remain costly and limited for some time, with or without the fault-tolerant capability. The \hbem model can also significantly reduce the efforts needed to optimize and to compile the QSVT algorithm. The \bem and its Hermitian version provides a simple and reliable way to generate random matrices on a quantum computer with the minimal use of the ancilla qubits and the controllability to the circuit depth. 
Therefore, we expect that the quantum LINPACK benchmark uses the minimal circuit to gauge the performance of a quantum computer for solving linear algebra problems.

Furthermore, we demonstrate that using \emph{the same quantum circuit but with different parameters}, the combination of QSVT and the \hbem model can be used to perform many other linear algebra tasks, such as computing spectral measures and performing time series analysis (without Trotter splitting). Using the minimally entangled typical thermal state (METTS) algorithm 
\cite{White2009,StoudenmireWhite2010,MottaSunTanEtAl2020}, we also show how \hbem simplifies the computation of the thermal average of the energy. These linear algebra tasks can also be used to construct benchmarks similar to the quantum LINPACK benchmark, whose performance reflects the minimal requirement for a quantum computer in solving corresponding scientific computing problems. We implement all algorithms on the IBM Q quantum architecture. Due to the high noise level of the currently available quantum architecture, we also demonstrate the numerical performance using quantum virtual machines (QVM) with tunable, approximate error models derived from quantum devices.

\section{Random circuit based block-encoding matrix}\label{sec:racbem}

\noindent\textit{Block-encoding:}
Inherently, quantum computers can only handle unitary operators. Hence any non-unitary operators must be encoded in terms of unitary operators. Let $A\in \CC^{N\times N}$ be an $n$-qubit Hermitian matrix ($N=2^n$). If we can find an $(n+1)$-qubit unitary matrix $U_A$ such that
\begin{equation}
U_A=\left(\begin{array}{cc}
{A} & {\cdot} \\
{\cdot} & {\cdot}
\end{array}\right)
\label{eqn:block_encode_exact_matrix}
\end{equation}
holds, i.e. $A$ is the upper-left matrix block of $U_A$, then we may get access to $A$ via the unitary matrix $U_A$ with $A=\left(\langle 0 | \otimes I_n\right) U_A \left( | 0 \rangle \otimes I_n \right)$. Clearly when the operator norm $\norm{A}_2$ is larger than $1$, $A$ cannot be encoded by any unitary $U_A$ as in \eqref{eqn:block_encode_exact_matrix}. Generally if we can find $\alpha, \epsilon \in \mathbb{R}_+$, and an $(m+n)$-qubit matrix $U_A$ such that
\begin{equation}
\Vert A - \alpha \left(\langle 0^m | \otimes I_n\right) U_A \left( | 0^m \rangle \otimes I_n \right) \Vert \leq \epsilon,
\label{eqn:block_encoding}
\end{equation}
then $U_A$ is called an $(\alpha, m, \epsilon)$-block-encoding of $A$. Here $m$ is called the number of ancilla qubits for block-encoding. The block-encoding is a powerful and versatile model, which can be used to efficiently encode density operators, Gram matrices, positive-operator valued measure (POVM), sparse-access matrices, as well as addition and multiplication of block-encoded matrices (we refer to \cite{GilyenSuLowEtAl2019} for a detailed illustration of such constructions).

The block-encoding model also has its limitation. Take an $n$-qubit, $d$-sparse matrix $A$ (i.e. the number of nonzero entries in each row/column does not exceed $d$) for example, assuming access to its row/column entries via certain sparse-access oracles, then we can construct a block-encoding using $n+3$ ancilla qubits \cite{GilyenSuLowEtAl2019}. Any further manipulation of $U_A$, such as quantum signal processing, would require using $(n+4)$-qubit Toffoli gates, which are relatively expensive to implement \cite{BarencoBennettCleveEtAl1995,SaeediPedram2013}. Another example is that $A$ is given by a linear combination of $K$ terms, each term being a tensor product of Pauli matrices. In the setting of the Hamiltonian simulation, $e^{\I At}$ can be simply approximated by exponentiating each term individually following a certain order via the Trotter-Suzuki formula. However, the block-encoding model would essentially require using linear combination of unitaries \cite{BerryChildsKothari2015}, which not only requires $\lceil \log K\rceil$ ancilla qubits, but also usage of $(\lceil \log K\rceil+1)$-qubit Toffoli gates to implement the prepare and select oracles needed to obtain the linear combination. Such operations are essentially forbidden for near-term applications due to  high error rates, and are still challenging when the number of qubits and the gate depth remain a limitation in reaching the desired accuracy on fault-tolerant devices.

\vspace{1em}
\noindent\textit{\bem:}
To harness the power of the block-encoding model and to avoid its pitfalls, we propose the Random circuit based block-encoding matrix (\bem) model as follows. Instead of first identifying $A$ and then finding its block-encoding $U_A$, we reverse this thought process: we first identify a unitary $U_A$ that is easy to implement on a quantum computer, and then ask which matrix can be block-encoded by $U_A$. 

\begin{figure}[htbp]
\begin{center}
\includegraphics[width=0.2\textwidth]{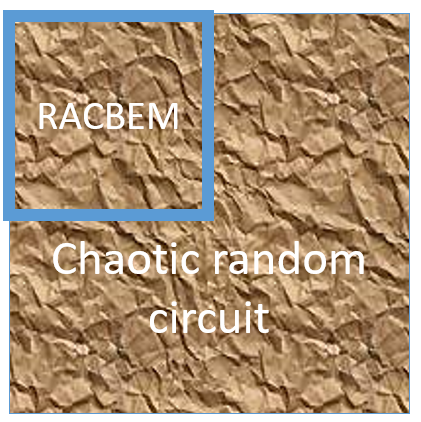}
\caption{A cartoon illustration of the \bem model.}
\end{center}
\end{figure}
It turns out that any matrix $A$ with $\norm{A}_2\le 1$ can be given by a $(1,1,0)$-block-encoding. Consider any $n$-qubit matrix with its singular value decomposition (SVD) $A=W\Sigma V^{\dag}$, where all singular values in $\Sigma$ belong to $[0,1]$. Then 
we may construct an $(n+1)$-qubit unitary matrix
\[
\begin{split}
U_A:=&\left(\begin{array}{cc}
A &  W \sqrt{I_{n}-\Sigma^{2}} \\
\sqrt{I_{n}-\Sigma^{2}} V^{\dagger} & -\Sigma
\end{array}\right)\\
=&\left(\begin{array}{cc}
W & 0 \\
0 & I_{n}
\end{array}\right)\left(\begin{array}{cc}
\Sigma & \sqrt{I_{n}-\Sigma^{2}} \\
\sqrt{I_{n}-\Sigma^{2}} & -\Sigma
\end{array}\right)\left(\begin{array}{cc}
V^{\dagger} & 0 \\
0 & I_{n}
\end{array}\right),
\end{split}
\]
which is a $(1,1,0)$-block-encoding of $A$. Since a random circuit with $\text{poly} (n)$ depth can approximate an $n$-qubit Haar measure at least according to the criterion of the $2$-design \cite{HarrowLow2009}, a sufficiently general $(n+1)$-qubit unitary $U_A$ can give access to in principle any $n$-qubit non-unitary matrices $A$ (up to a scaling factor). Furthermore, such a random circuit $U_A$ can be constructed using only basic one-qubit unitaries and CNOT gates. The matrix $A$ obtained by measuring the first qubit (or in fact, any qubit used as the ancilla) is called a \bem. Since the Haar measure is the uniform distribution of unitary matrices, we conclude that \bem is a proper generalization of dense matrices on quantum computers suitable for performing linear algebra tasks. The layout of the two-qubit operations can be designed to be compatible with the coupling map of the hardware.  

\begin{figure}[htbp]
    \centering
    \includegraphics[width=0.2\textwidth]{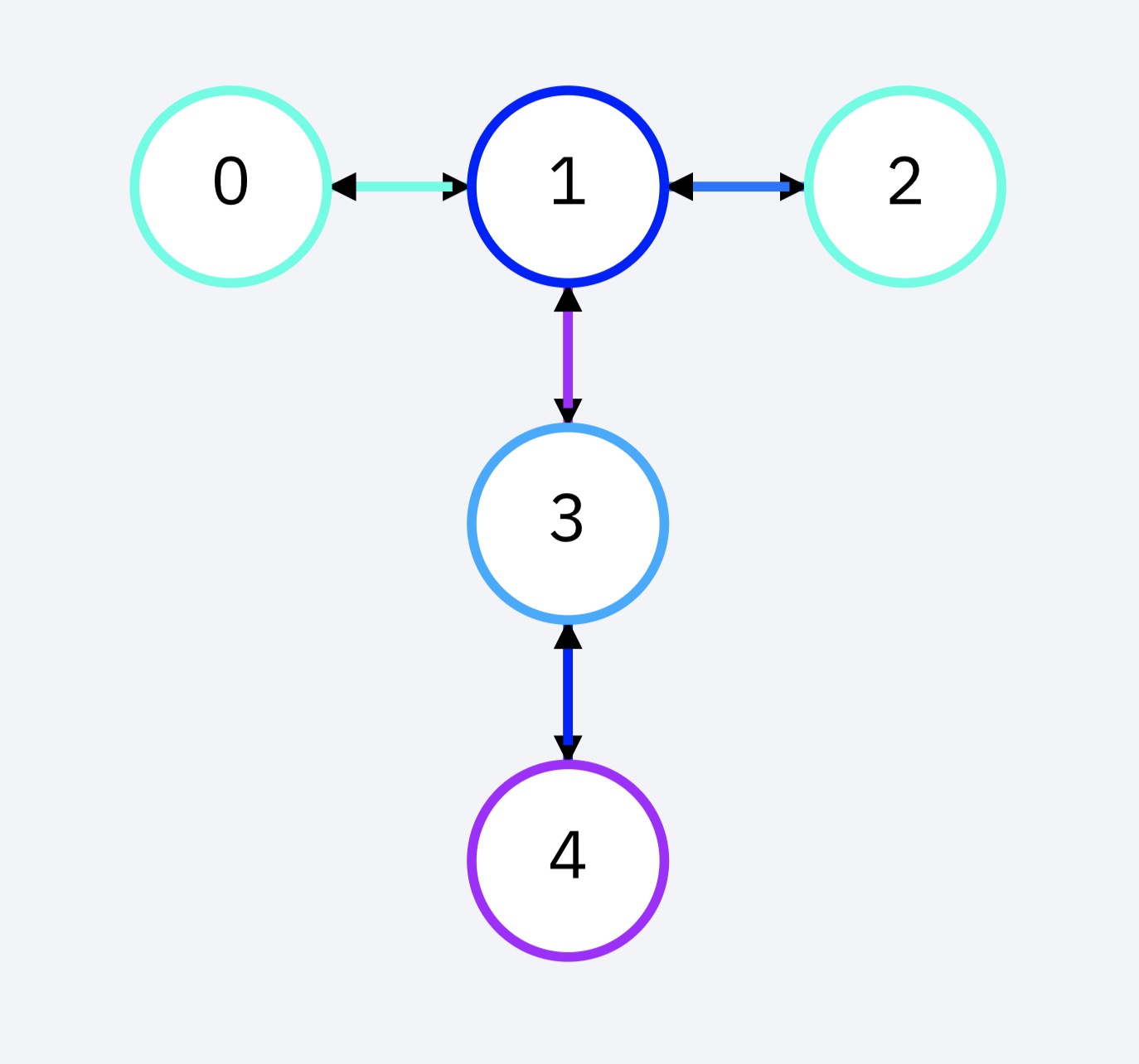}
    \caption{Coupling map of quantum computing backend \textsf{ibmq\_burlington} provided on IBM Q website \cite{IBMQ}. The encircled number stands for the label of qubit where the color represents the noise rate of one-qubit gates on it. If two qubits are related by an arrow, the CNOT gate is directly available between these two qubits.\label{fig:coupling_map}}
\end{figure}

For instance, for the \textsf{ibmq\_burlington} device with its coupling map shown in \cref{fig:coupling_map}, we can choose qubit 1 as the \bem ancilla qubit, which results in a \bem model with 3 system qubits $2,3,4$. The qubit 0 is not used here, and is reserved as the signal qubit for quantum singular value transformation to be discussed later.

\begin{figure*}[htbp]
    \centering
    \includegraphics[width=\textwidth]{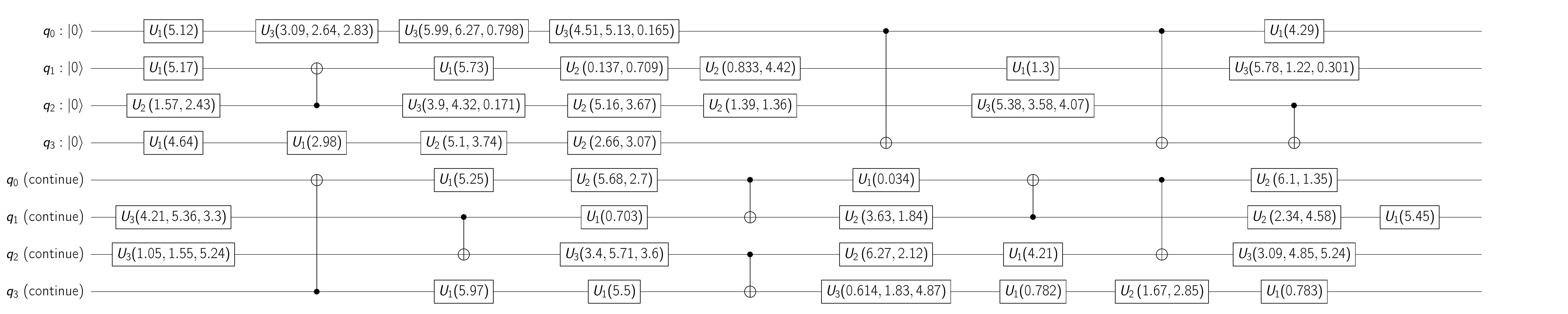}
    \caption{A \bem circuit constructed using the basic gate set $\{ \mathrm{U}_1, \mathrm{U}_2, \mathrm{U}_3, \mathrm{CNOT} \}$. The CNOT gates are directly implementable according to the coupling map in (a). $q_0, q_1, q_2, q_3$ refer to qubits $1,2,3,4$ in (a), where qubit $0$ is excluded as a signal qubit. The circuit at the bottom is a continuation of the top circuit.\label{fig:racbem_3qubit}}
\end{figure*}

\vspace{1em}
\noindent\textit{\hbem:}
In many applications, such as Hamiltonian simulation, thermal state preparation etc, we are only interested in Hermitian matrices.  It is possible to find a general circuit $U_A$ that coincidentally block-encodes a Hermitian matrix, but this can become increasingly difficult as $n$ increases. A useful fact is that once a random circuit $U_A$ is given, its Hermitian conjugate $U_A^{\dag}$ is easily accessible by conjugating the each gate and reversing the gate sequence. We will show below that this allows us to get access to in principle any $n$-qubit Hermitian matrix.

\begin{figure*}[htbp]
\begin{center}
\[
\Qcircuit @C=0.8em @R=1.em {
 \lstick{\ket{0}}& \gate{\mathrm{H}} & \targ & \gate{e^{-\I \varphi_0 \mathrm{Z}}} & \targ & \qw & \targ & \gate{e^{-\I \varphi_{1} \mathrm{Z}}} & \targ & \qw & \targ & \gate{e^{-\I \varphi_0 \mathrm{Z}}} & \targ & \gate{\mathrm{H}}&\qw  \\
\lstick{\ket{0}}& \qw  &\ctrlo{-1} & \qw  & \ctrlo{-1} & \multigate{1}{U_A} & \ctrlo{-1} & \qw & \ctrlo{-1} & \multigate{1}{U^{\dag}_A}    &\ctrlo{-1} & \qw & \ctrlo{-1} & \qw &\qw\\
\lstick{\ket{\psi}}& \qw &\qw &\qw&\qw &\ghost{U_A} &\qw&\qw&\qw&\ghost{U^{\dag}_A}&\qw&\qw   & \qw&\qw& \qw  
}
\]
\end{center}
\caption{Quantum circuit for generating a $(1,2,0)$-block-encoding of a \hbem from a $(1,1,0)$-block-encoding $U_A$ and its Hermitian conjugate. $\mathrm{H}$ is the Hadamard gate, and $\mathrm{Z}$ is the Pauli-Z gate.}
\label{fig:qsp_circuit_hermitian_racbem}
\end{figure*}
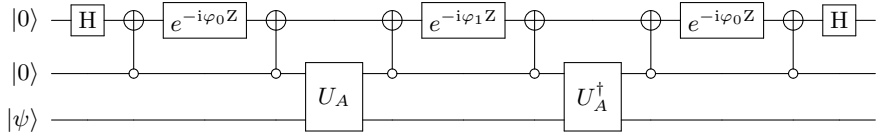 

Consider the quantum circuit in \cref{fig:qsp_circuit_hermitian_racbem} denoted by $U_{\mathfrak{H}}$, where $\varphi_0,\varphi_1\in [-\pi,\pi)$. Direct calculation (see \cref{sec:hracbem}) shows that
\begin{equation}
\begin{split}
\mathfrak{H}=&(\bra{0^2}\otimes I_n)U_{\mathfrak{H}}(\ket{0^2}\otimes I_n)\\
=&\left[-2 \sin (2\varphi_0) \sin \varphi_1\right] A^{\dag}A+\cos(2\varphi_0-\varphi_1).
\end{split}
\label{eqn:hracbem}
\end{equation}
Here $\mathfrak{H}$ is a Hermitian matrix. We refer to it as a Hermitian \bem (\hbem), and $U_\mathfrak{H}$ is its $(1,2,0)$-block-encoding.  In particular, choosing $\varphi_0=\pi/8$, $\varphi_1=-\pi/4$, then $\mathfrak{H}=A^{\dag} A$ is Hermitian positive semi-definite. This will be referred to as a \chbem.
In other words, a \chbem is constructed from its (non-unique) matrix square root $A$.

In \cref{fig:qsp_circuit_hermitian_racbem}, the CNOT gate controlling on $0$ instead of $1$ is mainly for notational convenience, and in fact not all CNOT gates are necessary here. For example, in order to implement a \chbem, we only need 1 application of $U_A$, 1 application of $U_A^{\dag}$, 2 H gates, 2 standard CNOT gates, 1 $\mathrm{S}^{\dag}$ gate, and 2 T gates (see \cref{sec:hracbem}). Since any matrix with singular values bounded by $1$ can be represented as a \bem, we immediately have that any Hermitian positive semi-definite matrix with eigenvalues bounded by $1$ can be represented as a \chbem, with a sufficiently flexible $U_A$.

\begin{figure*}[htbp]
\begin{center}
\[\scalebox{0.9}{
\Qcircuit @C=0.8em @R=1.em {
 \lstick{\ket{0}}& \gate{\mathrm{H}} & \targ & \gate{e^{-\I \varphi_{2d} \mathrm{Z}}} & \targ & \qw & \targ & \gate{e^{-\I \varphi_{2d-1} \mathrm{Z}}} & \targ & \qw & \qw &\raisebox{0em}{$\cdots$}&&\qw   &\targ & \gate{e^{-\I \varphi_0 \mathrm{Z}}} & \targ & \qw& \gate{\mathrm{H}}&\qw  \\
\lstick{\ket{0^m}}& \qw  &\ctrlo{-1} & \qw  & \ctrlo{-1} & \multigate{1}{U_A} & \ctrlo{-1} & \qw & \ctrlo{-1} & \multigate{1}{U^{\dag}_A} &\qw &\raisebox{0em}{$\cdots$} &&\qw    &\ctrlo{-1} & \qw & \ctrlo{-1} & \qw& \qw &\qw \\
\lstick{\ket{\psi}}& \qw &\qw &\qw&\qw &\ghost{U_A} &\qw&\qw&\qw&\ghost{U_A}&\qw &\raisebox{0em}{$\cdots$} &&\qw&\qw&\qw   & \qw&\qw& \qw &\qw  
}}
\]
\end{center}
\caption{Quantum circuit for quantum singular value transformation of a real matrix polynomial of degree $2d$. For QSVT of a \hbem, the number of ancilla qubits $m$ can be set to $1$.}
\label{fig:qsp_circuit_real}
\end{figure*}
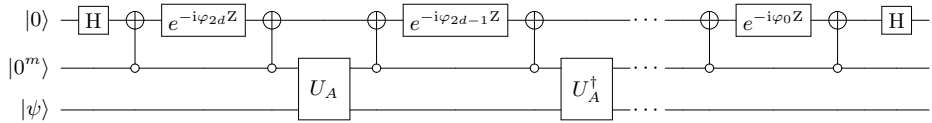

\vspace{1em}
\noindent\textit{Quantum singular value transformation:}
Given the SVD $A=W\Sigma V^{\dag}$, and a smooth function $f(x)$ of even parity, we define the quantum singular value transformation (QSVT) as 
\begin{equation}
f^{\triangleright}(A):=V f(\Sigma)V^{\dag}.
\label{eqn:right_generalize}
\end{equation}
Here the right pointing triangle reflects that only the right singular vectors $V$ are kept. Clearly $f^{\triangleright}(A)=f(\sqrt{A^{\dag}A})$, where the right hand side is the standard matrix function.  Now let $f$ be a real even polynomial of degree $2d$ that satisfies $\abs{f(x)}\le 1$ for any $x\in[-1,1]$. Let $U_A$ be a $(1,m,0)$-block-encoding of $A$. Then following \cite[Corollary 11]{GilyenSuLowEtAl2019}, there exists a $(1, m+1,0)$-block-encoding of $f^{\triangleright}(A)$, denoted by $U_{f^{\triangleright}(A)}$.  The circuit to implement $U_{f^{\triangleright}(A)}$ is given in \cref{fig:qsp_circuit_real}, which can be constructed  using $d$ queries of $U_A$ and $d$ queries of $U_A^{\dag}$.
Here $\Phi:=(\varphi_0,\ldots,\varphi_{2d})$ are called the phase factors. One challenge in QSVT is to find the phase factors $\Phi$. Besides the methods for obtaining $\Phi$ by polynomial factorization \cite{GilyenSuLowEtAl2019,Haah2019}, recently an optimization based method is proposed to find $\Phi$ up to very high degrees \cite{DongMengWhaleyEtAl}. A brief description of the method is given in \cref{sec:optimization}. 

Therefore \cref{fig:qsp_circuit_hermitian_racbem} implements a QSVT for a second order polynomial $\mathfrak{H}=h^{\triangleright}(A)$ with a symmetric choice of phase factors $\Phi=(\varphi_0,\varphi_1,\varphi_0)$, where
\begin{equation}
h(x)=\left[-2 \sin (2\varphi_0) \sin \varphi_1\right] x^2+\cos(2\varphi_0-\varphi_1).
\label{eqn:h_polynomial}
\end{equation}
A \chbem is given by $h(x)=x^2$. 

Consider any real polynomial $g(x)$ of degree $d$ without parity constraint, satisfying $\abs{g(x)}\le 1$ for any $x\in[-1,1]$.  Then using the identity
\begin{equation}
g(\mathfrak{H})=(g\circ h)^{\triangleright}(A):=f^{\triangleright}(A),
\label{eqn:poly_hracbem}
\end{equation}
any matrix function $g(\mathfrak{H})$ can be expressed as a QSVT with respect to an even polynomial $f=g\circ h$ of degree $2d$. We remark that when $g$ does not have a definite parity, the associated QSVT of $A$ is much more involved. It generally requires using a linear combination of block-encoding of the even and odd parts, which in turn requires implementing controlled $U_A$ \cite{GilyenSuLowEtAl2019}. Normally such controlled operations have significant overhead. For instance, if we would like to implement a controlled-\bem, generally we need to convert all quantum gates in the circuit of $U_A$ to the controlled version. 

By expressing $g(\mathfrak{H})$ as a QSVT associated with an even polynomial, we have not only eliminated the need of controlled unitary operations, but also saved one additional qubit. This is because if we first construct a \hbem $\mathfrak{H}$ and then construct $g(\mathfrak{H})$, we need $3$ ancilla qubits in total, and possibly a LCU circuit when $g$ does not have a definite parity. On the other hand, by considering the composite function $f$, the parity constraint on $g$ is completely removed, and we only need $2$ ancilla qubits (the same as that needed for a \hbem). Furthermore, each controlled gate in \cref{fig:qsp_circuit_real} is a standard CNOT gate rather than a Toffoli gate.  Therefore the \hbem model has a salient advantage and significantly simplifies the implementation.

In many applications, such as a LCU circuit and the Hadamard test, we do need to get access to controlled $U_{\mathfrak{H}}$ or $U_{g(\mathfrak{H})}$. In such a case, the fact that $\mathfrak{H}$ is a \hbem is also helpful: Note that if we remove all Z-rotations in the first line of \cref{fig:qsp_circuit_real}, the circuit implements an identity operator since $U_A,U_A^{\dag}$ always appear in pairs thanks to that $f=g\circ h$ is an even polynomial. Therefore a controlled version $U_{g(\mathfrak{H})}$ can be simply implemented by controlling all the Z-rotation gates \cite{GilyenSuLowEtAl2019}.

\section{Proposal of quantum LINPACK benchmark}\label{sec:linpack}

The \bem, as well as the \hbem model provides a solution to the read-in problem using only basic quantum gates, and we can design them to be optimally adapted to the hardware architecture without resorting to complex quantum compilers. Hence they can be regarded as the proper generalization of ``test dense matrices'' in the quantum setting.

In this section, we demonstrate that the usage of the \hbem model for solving QLSP. We assume $A=\mathfrak{H}$ is a \hbem, and without loss of generality we may take $\ket{b}=\ket{0^n}$.  Let the condition number of $\mathfrak{H}$ be denoted by $\kappa$, which is the ratio between the maximum and the minimum of the singular values of $\mathfrak{H}$. It is believed that the computational complexity for solving QLSP cannot generally be better than $\Or(\kappa^{1-\delta})$ for any $\delta>0$ \cite{HarrowHassidimLloyd2009}, and the cost of using QSVT to solve general linear systems scales as $\wt{\Or}(\kappa^2\log(1/\epsilon))$ \cite{GilyenSuLowEtAl2019}.
So the treatment of ill-conditioned matrices is very difficult especially on near-term devices. To reduce the circuit depth, in the near future it may be more productive to focus on solving well conditioned linear systems.  

Note that if $A$ has at least one singular value that is zero (or near zero), a \chbem $\mathfrak{H}$ is not invertible (or very ill-conditioned). Such events can occur more frequently as $n$ becomes large. It can be difficult to diagnose such a problem without first obtaining some spectral information of $A$, which is perhaps a more difficult task than solving the linear system problem itself. 

The \hbem model offers a new and natural way to solve this problem. From \cref{eqn:hracbem}, assuming $\cos(2\varphi_0-\varphi_1)>0$ and  $-2\sin(2\varphi_0)\sin\varphi_1>0$, and use that $0\preceq A^{\dag}A\preceq 1$, the condition number of $\mathfrak{H}$ can be bounded from above:
\[
\kappa(\mathfrak{H})\le \frac{\cos(2\varphi_0+\varphi_1)}{\cos(2\varphi_0-\varphi_1)}.
\]
Therefore the condition number of a \hbem is fully tunable by changing the phase factors $\varphi_0,\varphi_1$. According to \cref{eqn:h_polynomial}, this changes the second order polynomial function $h(x)$, so that $\mathfrak{H}:=h^{\triangleright}(A)\succ 0$ has a tunable, bounded condition number.

\begin{figure}[htbp]
    \centering
    \includegraphics[width=5cm]{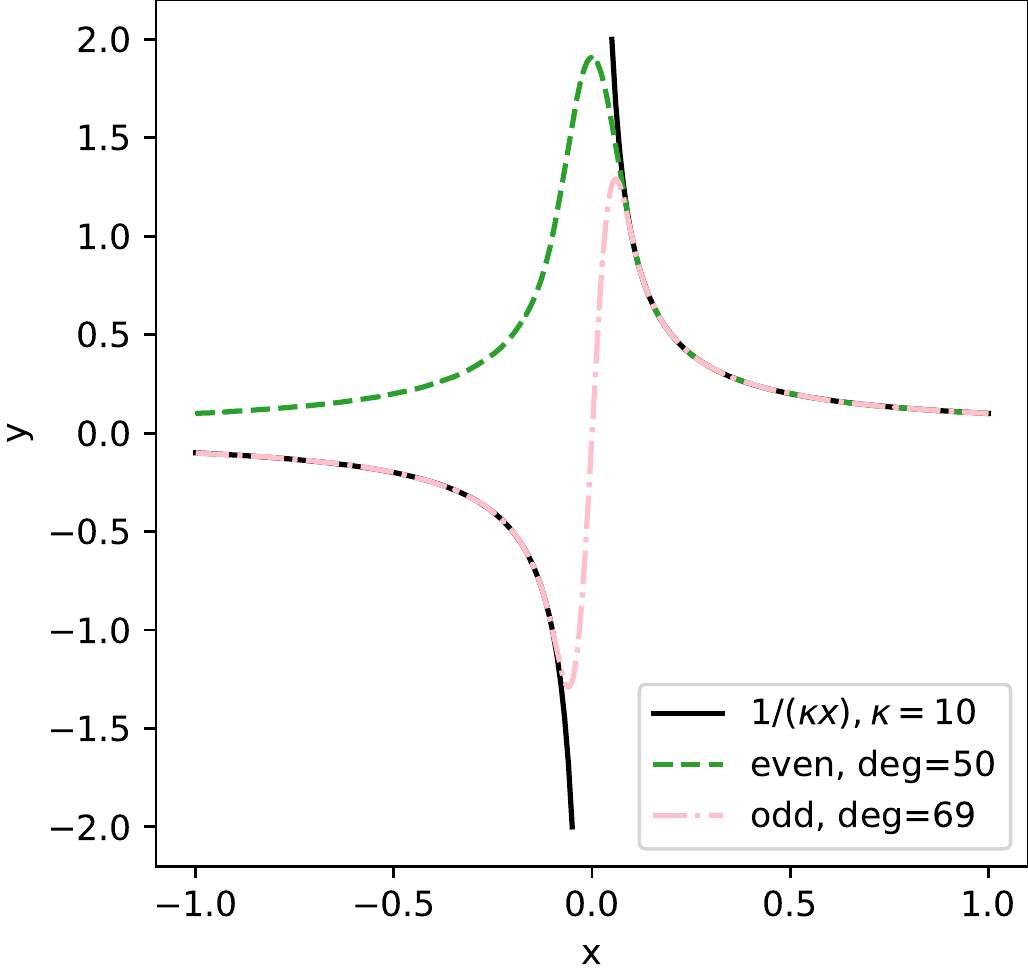}
    \caption{Approximate polynomials obtained by the Remez algorithm. The even and odd polynomials are generated such that the maximal approximation error to $1/(\kappa x)$ on $[\kappa^{-1}, 1]$ is $10^{-3}$. We find that the polynomial degree of the even approximation is lower than that of the odd approximation.}
    \label{fig:xinv}
\end{figure}

In order to solve QLSP, we are interested in finding a polynomial $g(x)$ of degree $d$ so that
\[
|g(x)-x^{-1}|\le \epsilon, \quad x\in[\kappa^{-1}, 1].
\]
Following \cite[Corollary 69]{,GilyenSuLowEtAl2018}, there can be satisfied by an odd polynomial $g(x)$ with degree $d\sim\Or(\kappa \log(1/\epsilon))$, which gives an upper bound on $d$. Numerical results shows that a better approximation to $g(x)$ can be obtained by solving a minimax problem using the Remez algorithm \cite{DongMengWhaleyEtAl}, and the polynomial can be chosen to be either even or odd.  \cref{fig:xinv} shows the shape of the optimal polynomial even/odd approximation to $x^{-1}$ on the interval $[\kappa^{-1},1]$ found by the Remez algorithm to reach a target accuracy of $10^{-3}$ with $\kappa=10$. In particular, the usage of an even polynomial can further reduce the polynomial degree, which will be used in the numerical tests below.

We may then construct a degree $2d$ polynomial $f=g\circ h$ as in \cref{sec:racbem}. Once we find the associated phase factors, the circuit in \cref{fig:qsp_circuit_real} implements $g(\mathfrak{H})\ket{b}$, which satisfies the error bound $\norm{g(\mathfrak{H})\ket{b}-\kappa^{-1} \mathfrak{H}^{-1}\ket{b}}_2\le \epsilon$ for any normalized vector $\ket{b}$. The success probability of measuring the ancilla qubits and obtaining an all $0$ output (will be referred to as the success probability for short) is
\[
p=\norm{g(\mathfrak{H})\ket{b}}_2^2\ge (\kappa^{-1}-\epsilon)^2,
\]
which can be of modest size given $\kappa$ is not too large.

Since measuring the accuracy of all entries of the solution is not a practical goal, our proposal of the quantum LINPACK benchmark is to \emph{measure the success probability $p$}, i.e. the probability of measuring the $2$ ancilla qubits and obtaining $\ket{00}$, and to compare it with the numerically exact probability (denoted by $p_{\mathrm{exact}}$) computed using a classical computer. When  $\kappa,\epsilon$ and \hbem are given, the quantum LINPACK benchmark reports the relative error $\abs{p-p_{\mathrm{exact}}}/p_{\mathrm{exact}}$ to describe the accuracy of a quantum computer for solving QLSP. In the future we may also take into account the wall clock time, or a quantity analogous to the floating point operations per second (FLOPS) for classical computers to measure the efficiency of a quantum computer. The quantum LINPACK benchmark uses only basic single-qubit gates and CNOT gates, and hence is easy to implement even on near-term devices. 

Clearly, the success of the quantum LINPACK benchmark is only a \emph{necessary condition} for the accurate solution of QLSP. However, as will be shown in numerical experiments, this task can already be challenging for near-term devices due to the presence of noise, and therefore the benchmark provides a meaningful and easily implementable criterion for measuring the accuracy of quantum computers. Furthermore, due to the very flexible construction of $U_A$ and the near-minimal usage ancilla qubits and other primitive gates, we also expect that the success of quantum LINPACK benchmark is the \emph{minimal requirement} in order to achieve quantum advantages by solving QLSP or similar linear algebra tasks.

\section{Other quantum linear algebra applications}

Since QSVT implements a general matrix function, its application is therefore not limited to solving QLSP. Here we demonstrate a few more applications. Throughout the section $\mathfrak{H}$ is a \hbem defined via a unitary matrix $U_A$ and $h(x)$ is a second order polynomial. 

\vspace{1em}
\noindent\textit{Time series analysis:} Given an $n$-qubit state $\ket{\psi}$, consider the computation of the following time series
\begin{equation}
s(t)=\braket{\psi|e^{\I \mathfrak{H}t}|\psi}.
\label{eqn:signal_proc}
\end{equation}
When $\mathfrak{H}$ is a \hbem, we can evaluate $s(t)$ by measuring the its real and imaginary components separately:
\begin{equation}
\begin{aligned}
s(t)=&\braket{\psi|\cos (\mathfrak{H}t)|\psi}+\I\braket{\psi|\sin (\mathfrak{H}t)|\psi}\\
=&2\Braket{\psi|\frac12(\cos(\mathfrak{H}t)+I_n)|\psi}\\
&+2\I\Braket{\psi|\frac12(\sin(\mathfrak{H}t)+I_{n})|\psi}-(1+\I)\\
=&:2\norm{g_{c,t}(\mathfrak{H})\ket{\psi}}_2^2+2\I\norm{g_{s,t}(\mathfrak{H})\ket{\psi}}_2^2-(1+\I).
\end{aligned}
\label{eqn:signal_proc_identity}
\end{equation}
Here we introduce the functions for the cosine and sine part, respectively
\[
g_{c,t}(x)=\sqrt{\frac{\cos(xt)+1}{2}}, \quad 
g_{s,t}(x)=\sqrt{\frac{\sin(xt)+1}{2}}.
\]
Now the quantities in \cref{eqn:signal_proc_identity} can be directly obtained via the success probability of the QSVT circuit with $f_{c,t}:=g_{c,t}\circ h$ and $f_{s,t}:=g_{s,t}\circ h$, respectively (with a suitable polynomial approximation of $g_{c,t},g_{s,t}$). Note that the access to the matrix square root of $\mathfrak{H}$ here is crucial: though $g_{c,t}$ is an even function itself, $g_{s,t}$ does not have a definite parity, and therefore the direct implementation of $g_{s,t}(\mathfrak{H})$ would require a LCU circuit and hence controlled unitary operations. In the setting of \hbem, the treatment of $g_{c,t}$ and $g_{s,t}$ can be put on the same footing due to the composition with the even polynomial $h$. 

In order to accelerate the convergence of the polynomial approximation, in practice, we may introduce another tunable parameter $\eta \ge 1$ in the formulation so that 
\[
g_{c,t,\eta}(x) = \sqrt{\frac{\cos(xt)+\eta}{2}}, \quad 
g_{s,t,\eta}(x) = \sqrt{\frac{\sin(xt)+\eta}{2}}
\] 
and then
\[
s(t) = 2\norm{g_{c,t,\eta}(\mathfrak{H})\ket{\psi}}_2^2 + 2\I\norm{g_{s,t,\eta}(\mathfrak{H})\ket{\psi}}_2^2 - \eta(1+\I).
\]
For instance, $\eta$ can be set to $1.0$ or $1.5$ and the details are given in \cref{tab:param-ts}.

In \cite{Somma2019}, the time series analysis is used for eigenvalue estimation, which is implemented via the Hadamard test and requires a controlled Hamiltonian evolution procedure, followed by a Fourier transform. The procedure above removes the need of performing controlled Hamiltonian evolution. We also remark that in terms of quantum estimation of eigenvalues, the method using the spectral measures to be discussed below can be more advantageous, which does not require a subsequent Fourier transform, and the resulting spectral measure is guaranteed to be non-negative. 

\vspace{1em}
\noindent\textit{Spectral measure:} Given an $n$-qubit state $\ket{\psi}$, in order to approximately evaluate the spectral measure, we may use the Plemelj formula
\begin{equation}
\begin{split}
s(E)=&\braket{\psi|\delta(\mathfrak{H}-E)|\psi}\\
=&\lim_{\eta\to 0^+} \frac{1}{\pi} \Im \braket{\psi|(\mathfrak{H}-E-\I \eta)^{-1}|\psi}, \quad E\in[-1, 1].
\end{split}
\label{eqn:spec_measure}
\end{equation}
By choosing a finite $\eta$ as the broadening parameter, we need to evaluate \[
s_{\eta}(E)=\frac{\eta}{\pi}\Braket{\psi|((\mathfrak{H}-E)^2+\eta^2)^{-1}|\psi}.
\]
Now define 
\[
g_{\eta,E}(x)=\left(\frac{\eta^2}{(x-E)^2+\eta^2}\right)^{\frac12},
\]
which satisfies $|g_{\eta,E}|\le 1$ for $x\in[-1,1]$, then $g_{\eta,E}\circ h$ can be approximated by an even order polynomial. Therefore 
\[
s_{\eta}(E)=\frac{1}{\eta\pi}\norm{g_{\eta,E}(\mathfrak{H})\ket{\psi}}^2_2,
\] 
and it can be obtained via the success probability of the QSVT circuit with $f=g_{\eta,E}\circ h$ (with a suitable polynomial approximation of $g_{\eta,E}$).

\vspace{1em}
\noindent\textit{Thermal averages:} In order to evaluate the thermal average of the energy ($\beta$ is the inverse temperature)
\[
E(\beta)=\frac{\Tr[\mathfrak{H} e^{-\beta \mathfrak{H}}]}{\Tr[e^{-\beta \mathfrak{H}}]},
\]
we may use the  minimally entangled typical thermal state (METTS) algorithm \cite{White2009,StoudenmireWhite2010,MottaSunTanEtAl2020}. Let $Z=\Tr[e^{-\beta \mathfrak{H}}]$, and
\[
E(\beta)=\frac{1}{Z}\sum_{i\in[N]}\braket{i|e^{-\beta \mathfrak{H}/2} \mathfrak{H} e^{-\beta \mathfrak{H}/2}|i}=\sum_{i\in[N]} p_{i}\braket{\phi_i|\mathfrak{H}|\phi_i}.
\]
Here $i$ loops over all states of the computational basis, each represented by an $n$-bit string. From the unnormalized states $\ket{\wt{\phi}_i}= e^{-\beta \mathfrak{H}/2}\ket{i}$, we define a probability distribution  $p_i=\braket{\wt{\phi_i}|\wt{\phi_i}}/Z$, and corresponding normalized states
$\ket{\phi_i}=\ket{\wt{\phi}_i}/\sqrt{\braket{\wt{\phi_i}|\wt{\phi_i}}}$. For each $i$, the contribution to the energy can be expressed as
\begin{equation}
\braket{\phi_i|\mathfrak{H}|\phi_i}=\frac{\braket{i|e^{-\beta \mathfrak{H}} \mathfrak{H} |i}}{\braket{i| e^{-\beta \mathfrak{H}}|i}}
=\frac{\norm{g_{n,\beta}(\mathfrak{H})\ket{i}}_2^2}{\norm{g_{d,\beta}(\mathfrak{H})\ket{i}}_2^2}.
\label{eqn:thermal_eachterm}
\end{equation}
Here we define
\[
g_{n,\beta}(x)=e^{-\beta x/2} \sqrt{x}, \quad g_{d,\beta}(x)=e^{-\beta x/2},
\]
for the numerator and the denominator, respectively, and without loss of generality we assume $\mathfrak{H}\succeq 0$. The numerator and the denominator can be obtained via the success probability of the QSVT circuit with $f_{n,\beta}:=g_{n,\beta}\circ h$ and $f_{d,\beta}:=g_{d,\beta}\circ h$, respectively (with a suitable polynomial approximation of $g_{n,\beta},g_{d,\beta}$).   

Unlike Metropolis type algorithms which follows an acceptance / rejection procedure, the METTS algorithm samples the states $\{\ket{\phi_i}\}$ as follows. We start from  a computational basis state $\ket{i}$ (e.g. $\ket{i}=\ket{0^n}$). Then
\begin{enumerate}
\item Evaluate the contribution to the energy from the state $\ket{\phi_i}$ via \cref{eqn:thermal_eachterm}.

\item Collapse $\ket{\phi_i}$ to a new state in the computational basis $\ket{i'}$ with probability $|\braket{i'|\phi_i}|^2$, and repeat step 1.
\end{enumerate}

Note that step $2$ of the METTS algorithm is very simple to implement: we only need to construct a QSVT circuit for preparing the unnormalized state $\ket{\wt{\phi}_i}$, which is readily available when computing the denominator in \cref{eqn:thermal_eachterm}. Then collapsing to $\ket{i'}$ can be implemented by measuring all $n$ system qubits in the computational basis. 

The evaluation of $f_{n,\beta} := g_{n,\beta}\circ h$ requires approximating the square root function. When $\mathfrak{H}$ is a \chbem with $h(x) = x^2$, an alternative method is to approximate an odd function $\wt{f}_{n,\beta}(x) = x e^{-\beta x^2 / 2}$ instead of $f_{n,\beta}$. Then the complexity for thermal average calculation can be only $\wt{\Or}(\sqrt{\beta}\log(1/\epsilon))$ \cite[Corollary 64]{GilyenSuLowEtAl2018}.

\section{Numerical results}
The source code of \bem is available in the Github repository\footnote{\url{https://github.com/qsppack/RACBEM}}. We use the optimization-based algorithm which is described in \cref{sec:optimization} to generate phase factors and the source code is available in the Github repository\footnote{\url{https://github.com/qsppack/QSPPACK}}. We demonstrate the numerical performance of the \bem  model for solving various numerical quantum linear algebra tasks. The algorithms are performed on the IBM Q quantum device, as well as QVM with approximate noisy running environment retrieved from quantum devices. All numerical tests are implemented in \textsf{python3.7} and \textsf{Qiskit} \cite{Qiskit}. In order to construct an adjustable noise model, we retrieve the noise model from real quantum devices provided by IBM Q backends. The magnitude of the noise level is then made to be fully tunable through a single parameter $\sigma$. When $\sigma=0$, the only noise contribution comes from the Monte Carlo error in measurements, which can be systematically reduced by increasing the number of samples. When $\sigma=1$, the noise model contains all the readout errors and quantum errors associated with a probability distribution given by the retrieved noise model (see \cref{app:model} for details). Unless otherwise noted, the number of measurements (shots) is fixed to be $8192$ throughout this section.  As the quantum linear algebra tasks are solved via QSVT, the overall error consists of contributions from the polynomial approximation, the noise from the device, and the Monte Carlo sampling error. We remark that in numerical results we distinguish the setup of ``QSVT without error'' (only taking into account the polynomial approximation error) from that of ``$\sigma=0$'' (including both the Monte Carlo sampling error and the polynomial approximation error).

We use \cref{alg:custom-rqc} (in \cref{app:model}) to generate custom random quantum circuits with respect to a given coupling map. In each layer of the  quantum circuit, we apply a one-qubit (or two-qubit) gate to each qubit (or a pair of qubits) randomly selected from the basic gate set. Though CNOT gates can increase the entanglement among the qubits, their error rate is much higher than that of one qubit gates on IBM Q backends. So we control the number of CNOT gates via a parameter to determine the probability that a CNOT gate is drawn. Unless otherwise noted, this probability is set to be $p = 0.5$. The circuit depth is the same as the number of layers in the generating circuit. We would like to emphasize that the circuit depth must not be too small. Otherwise the block-encoded matrix can sometimes become degenerate (such as a scaled identity matrix). For an $n$-qubit system, we empirically set the depth $\ell$ to be $\ell = 3$ when $n = 1$, $\ell = 7$ when $n = 2$, and $\ell = 15 + 2(n-3)$ when $n \ge 3$. We report the statistics of singular-value distributions of the block-encoded matrix in \cref{fig:svd-dist} (\cref{app:details}) to justify this adaptive choice of the circuit depth.

According to the random quantum circuit generation algorithm, we measure the total gate count of a quantum circuit by its logical gate count with respect to the basic gate set, which is upper bounded by its depth $\ell$ times the number of system qubits $n$. The available basic gates provided by IBM Q backends are the CNOT gate and U1, U2, U3 gates, which are parameterized families of generic one-qubit gates (see \cref{app:model} for details). In order to reduce the noise due to  U3 gates, we restrict the basic gate set in the custom random circuit generator to be \{CNOT, U1, U2\}, which is still a universal gate set. Each controlled rotation in QSVT circuit costs $7$ logical gates (4 X gates implemented by U2 gates, 2 CNOT gates, and 1 U1 gate). Therefore, given a \bem whose depth is $\ell$ and with $n$ system qubits, to implement the QSVT of a real polynomial of degree $\qspdeg$, the total gate count for the circuit is upper bounded by $2 + 7(\qspdeg+1) + \qspdeg  \ell n$.  The details about QSVT phase factors used in numerical experiments can be found in \cref{app:details}. Unless otherwise noted, the input quantum state of the quantum circuit is set to $\ket{0^{n+2}}$.

\begin{figure}[htbp]
    \centering
    \includegraphics[width = 8cm]{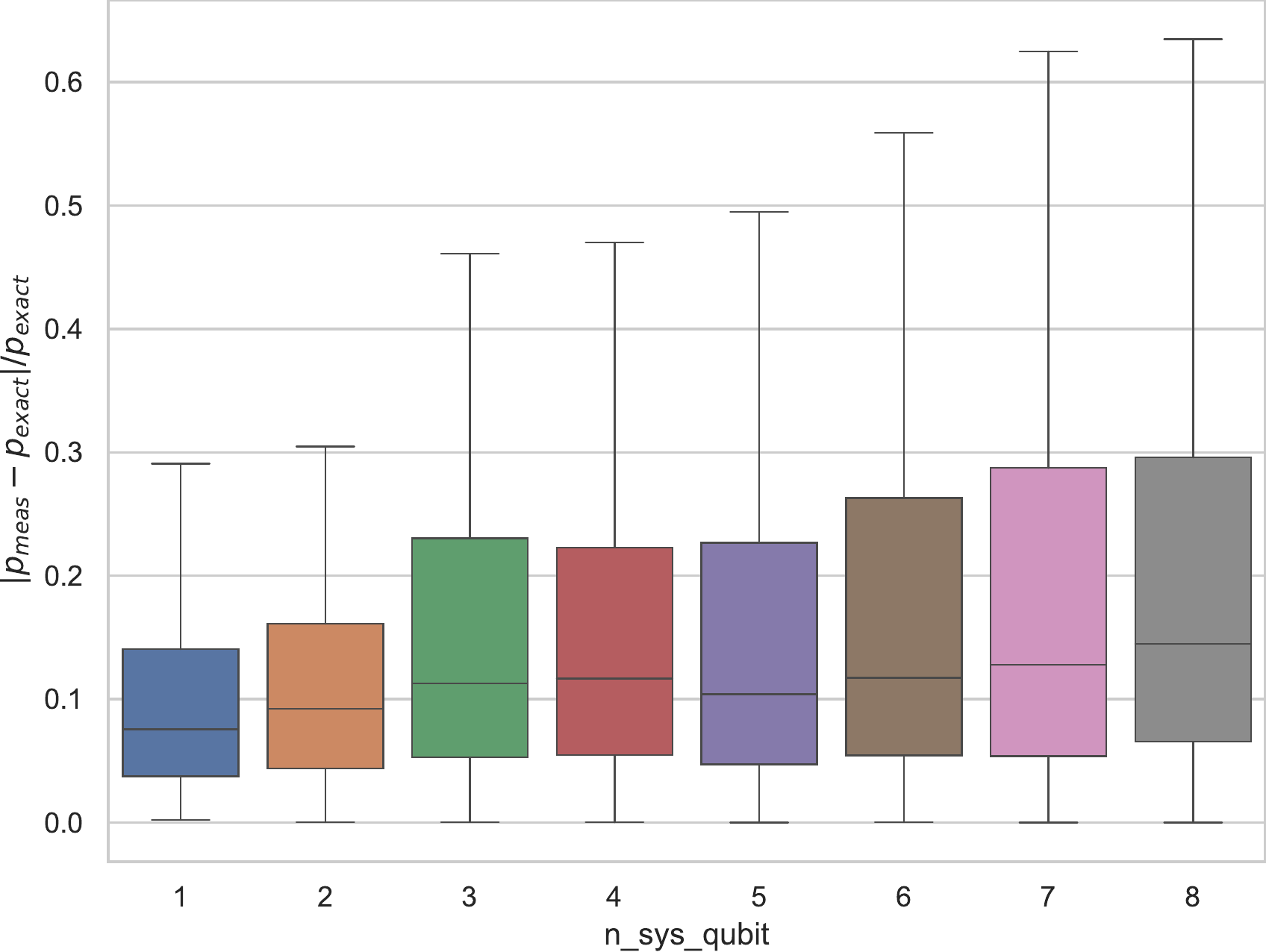}
    \caption{Relative error in success probability of obtaining $A\ket{0^n}$ for different number of system qubits.  The box ranges from the first to the third quartiles (25\% and 75\% percentiles respectively) of the dataset. The horizontal line inside the box stands for the median of the dataset. The whisker extending from the box indicates the rest of the dataset. The \bem's are generated using our custom random circuit generator in which the coupling map, basic gates and logical-to-physical layout are specified. When the number of system qubits is less than or equal to $3$, the results are obtained on the $5$-qubit backend \textsf{ibmq\_burlington}. Other results are obtained using $15$-qubit backend \textsf{ibmq\_16\_melbourne}. The dataset consists of results from $200$ \bem's when $n=1,2,3$, $300$ \bem's when $n=4,5,6$, $400$ \bem's when $n=7$ and $500$ \bem's when $n=8$.}
\label{fig:be-ibmq-benchmark}
\end{figure}

\vspace{1em}
\noindent\textit{\bem:}
Before presenting results of various numerical linear algebra problems, we measure the effect of noise on \bem directly on quantum computing backends provided by IBM Q. The numerical results are displayed in \cref{fig:be-ibmq-benchmark}. Each data point is obtained by generating a \bem using \cref{alg:custom-rqc}, and we measure the success probability of obtaining the block-encoded matrix applied to $\ket{0^n}$, which is equal to $\norm{A\ket{0^n}}_2^2$. The number of repeated measurements (shots) is 8192. \cref{fig:be-ibmq-benchmark} shows that the relative error of the quantum device can be considerable, ranging  from $10\%$ to $30\%$ as $n$ increases. Since the \bem is the building block of all subsequent quantum linear algebra tasks, we expect that the relative error of such tasks on quantum computing backends provided by IBM Q should be at least around the same level. 

\vspace{1em}
\noindent\textit{QLSP:} According to the discussion in \cref{sec:linpack}, it is possible to measure the accuracy of the solution to QLSP by sampling the output distribution of $g(\mathfrak{H})\ket{0^n}$ using e.g. a cross-entropy test similar to that in \cite{AruteAryaBabbushEtAl2019}. Here for simplicity we only measure $\norm{g(\mathfrak{H})\ket{0^n}}_2^2$ as a success probability (of measuring all ancilla qubits and obtain 0's), and evaluate the relative error compared to the exact value $\norm{\scalefac^{-1} \mathfrak{H}^{-1}\ket{0^n}}_2^2$ evaluated on a classical computer where $\scalefac$ is a scaling factor (see \cref{tab:param-qlsp} for details). The small relative error in success probability is a necessary condition to ensure the solution is correct which yields the benchmark in a weak sense. The condition number of \hbem is controlled by a second order polynomial $h_\kappa(x) := (1-\kappa^{-1})x^2 + \kappa^{-1}$ as in \cref{sec:linpack}. Another polynomial $g_\kappa(x)$ which approximates $x^{-1}$ is chosen to perform matrix inversion. The composite polynomial $g_\kappa \circ h_\kappa$ can be implemented by an even order QSVT circuit to carry out the overall procedure with only two ancilla qubits.

\begin{figure}[htbp]
    \centering
    \includegraphics[width=8cm]{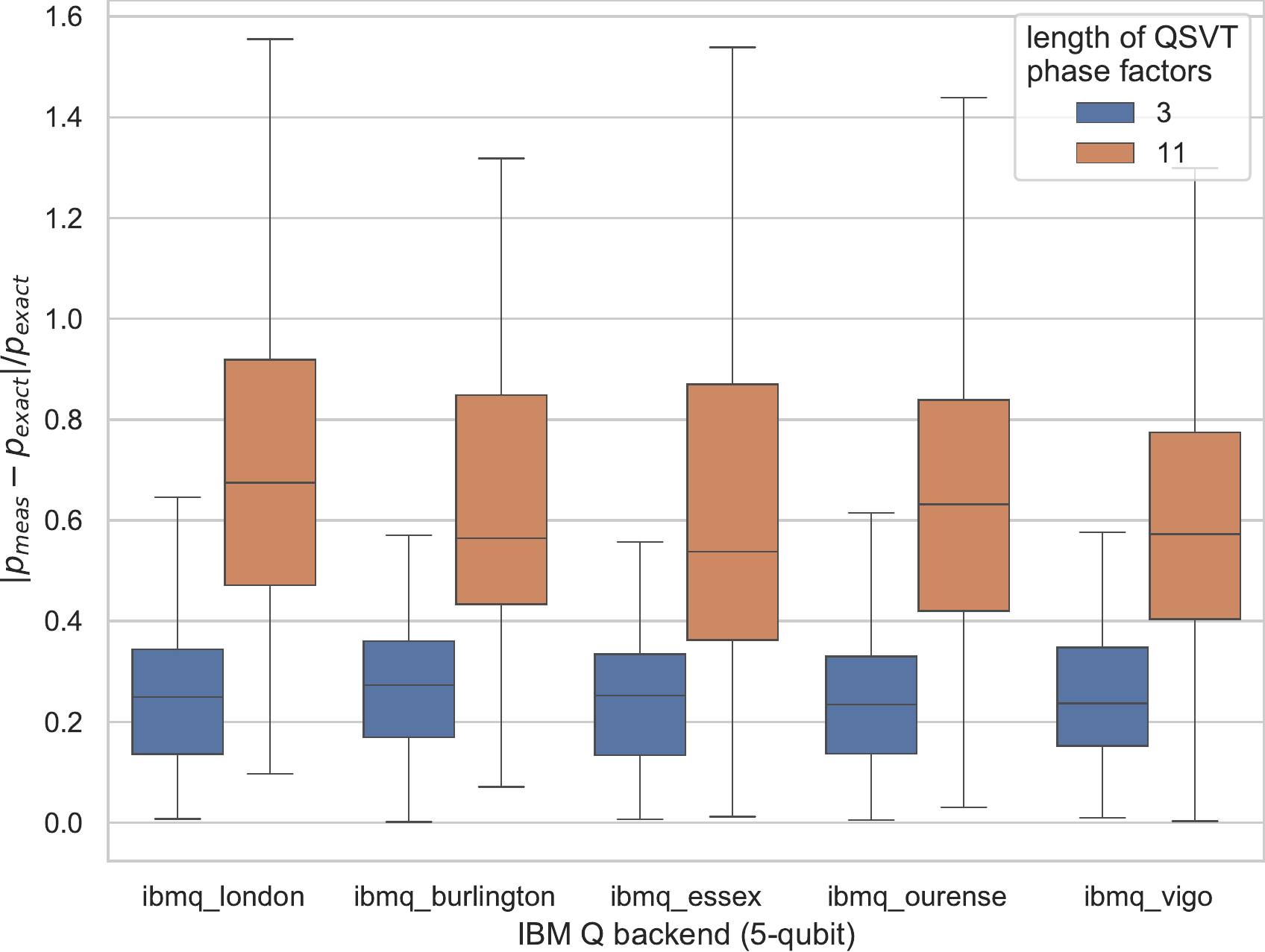}
    \caption{Relative error of solving QLSP using the \hbem model on IBM Q 5-qubit backends with identical quantum architectures. The condition number of each \hbem is upper bounded by $2$ and the number of system qubits is $3$. The size of the box indicates the first and the third quartiles and the extending whisker indicates the rest of the dataset. The phase factor sequences of length $3$ and $11$ are used to carry out QSVT to solve QLSP. Each dataset is obtained using $100$ samples at random. The details about phase factors are given in \cref{tab:param-qlsp}.}
    \label{fig:ls-ibmq-benchmark}
\end{figure}

We report the performance of solving linear systems on the IBM Q device using the \hbem model in \cref{fig:ls-ibmq-benchmark}. The architecture of the five backends are identical, and therefore we may draw a \hbem at random and test it on all five backends. By tuning $h(x)$, the condition number of $\mathfrak{H}$ is upper bounded by $2$. Therefore we may even use a very short QSVT circuit with $d=2$, and the corresponding number of phase factors is $3$. This is essentially a linear approximation to the inverse, and the accuracy measured by the $L^\infty$ error is less than $0.03$. In such a case, the total logical gate count is upper bounded by $113$. We can refine the polynomial approximation by using a more accurate, and deeper QSVT circuit with $d=10$ (the number of phase factors is $11$, and the $L^\infty$ error is less than $3\times 10^{-5}$). In a case when $d=10$, the total logical gate count is upper bounded by $529$. The results in \cref{fig:ls-ibmq-benchmark} indicate that for the shallow circuit,  the relative error of the success probability is similar to that observed in \cref{fig:be-ibmq-benchmark}, which only measures the success probability of the block-encoding. However, the relative error is significantly larger using the deeper circuit, despite that the QSVT circuit implements a more accurate polynomial approximation to the matrix inverse. Thus, we conclude that when designing the quantum circuit, a proper choice of QSVT phase factors is needed which reflects the tradeoff between the polynomial approximation error and the error caused by the noisy running environment.

\begin{figure}[htbp]
    \centering
    \includegraphics[width = 8cm]{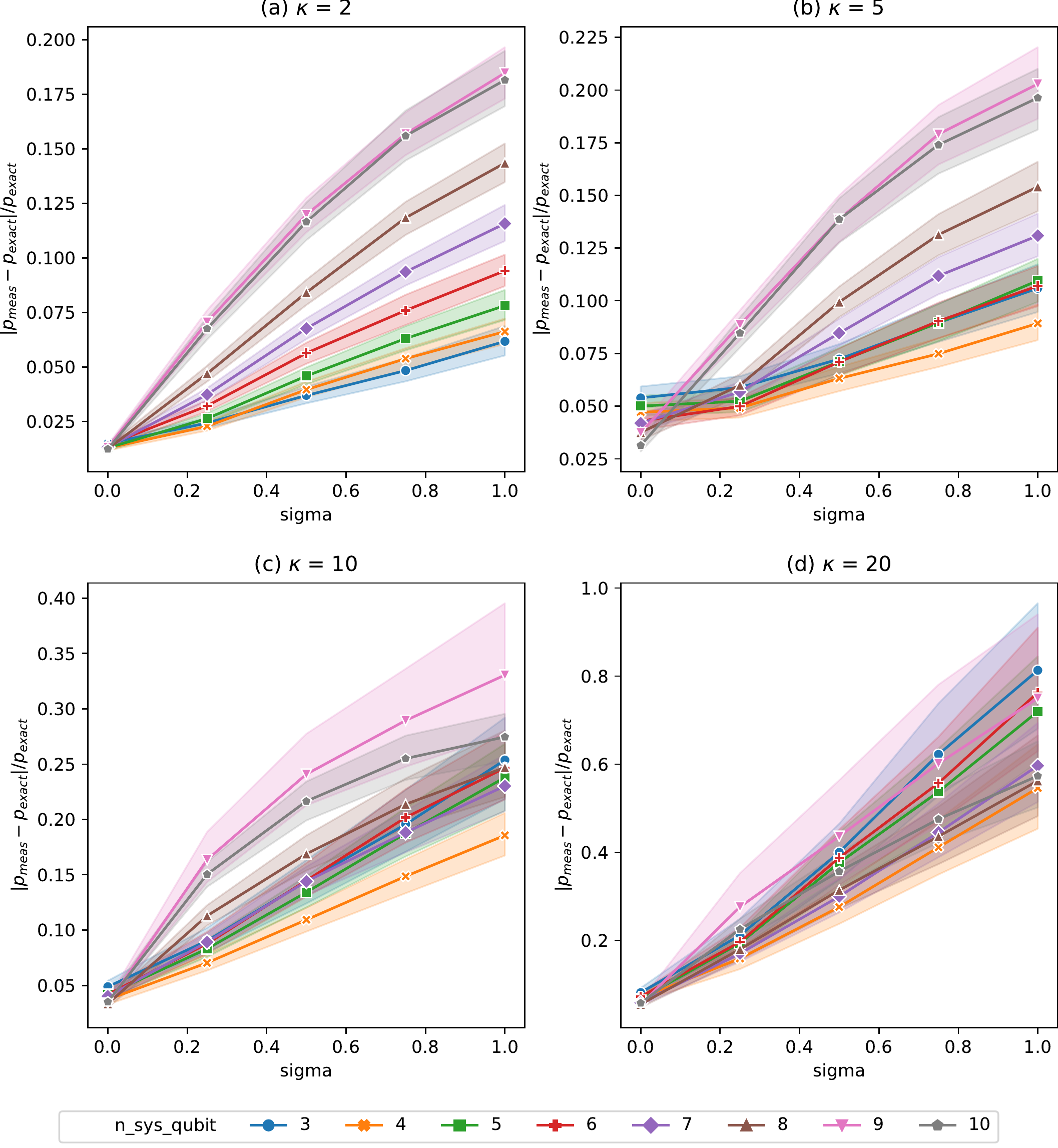}
    \caption{Relative error of solving QLSP for different number of system qubits, condition number and noise magnitude. The numerical results are obtained by running on QVM with architecture and noise model retrieved from IBM Q backend \textsf{ibmq\_16\_melbourne}. Each kind of parameters is computed by solving QLSP of $300$ \bem's at random. The marker is the average of each dataset and the shaded area is the 95\% confidence interval inferred from the dataset. The details about phase factors are given in \cref{tab:param-qlsp}.}
    \label{fig:ls-qasm-kappa}
\end{figure}

In order to further demonstrate how various parameters can affect the accuracy of the QLSP solver, we vary the number of system qubits, the condition number and the noise magnitude, and compute the relative error of success probability under these different settings on QVM. The numerical results are presented in \cref{fig:ls-qasm-kappa}. In all cases, we find that the QLSP solver can perform very well when $\sigma$ is small. The error is only due to the polynomial approximation and the Monte Carlo sampling error. This corresponds to the setting of fault-tolerant quantum devices. However, the accuracy rapidly deteriorates as  $\sigma$ increases. In the noisy setup, the error also increases nearly proportionally to the condition number $\kappa$. This is because the polynomial degree $d$ should increase as $\Or(\kappa)$ in order to achieve constant accuracy, and so is the circuit depth.  The error also increases with respect to the number of system qubits, but the effect is less significant compared to that due to $\kappa$ which increases the circuit depth.

\begin{figure*}[htbp]
    \centering
    \vspace{-10pt}
    \subfigure[\label{fig:ts-ibmq}]{
        \includegraphics[width=11cm]{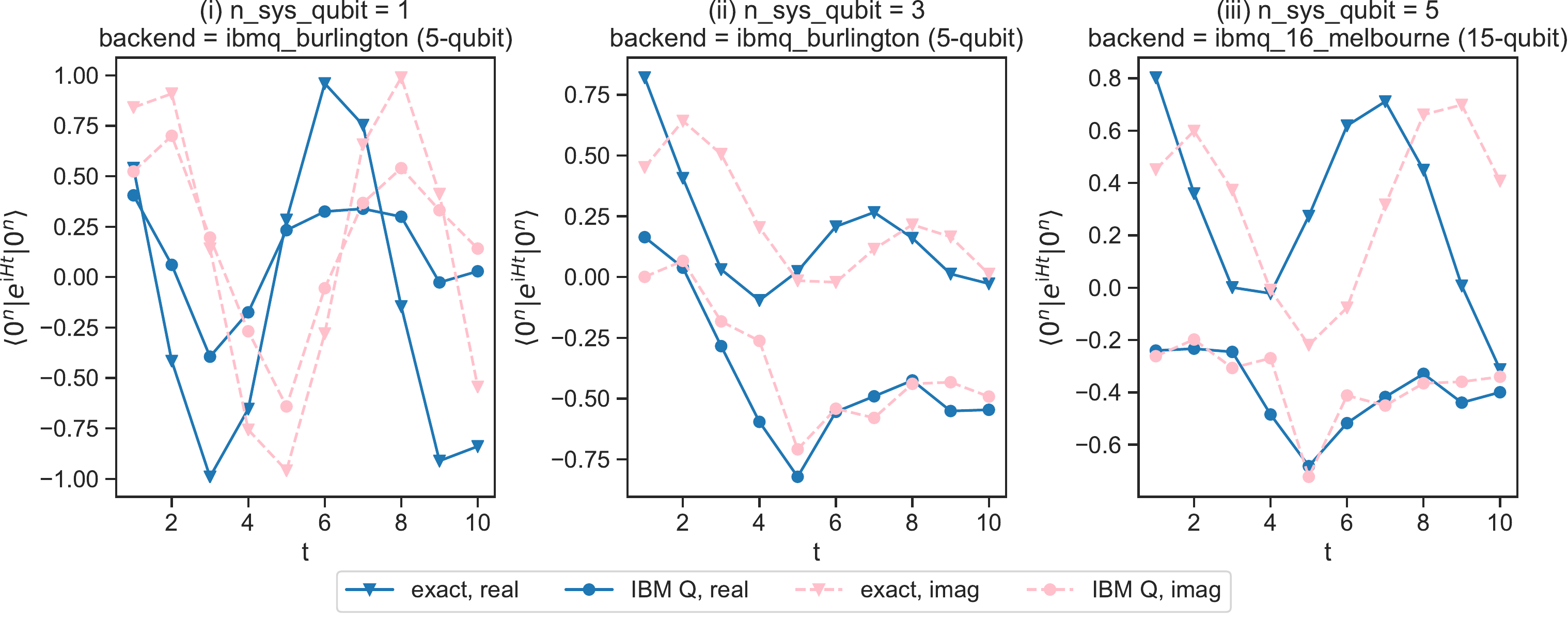}
    }
    
    \subfigure[\label{fig:ts-qasm}]{
        \includegraphics[width=14cm]{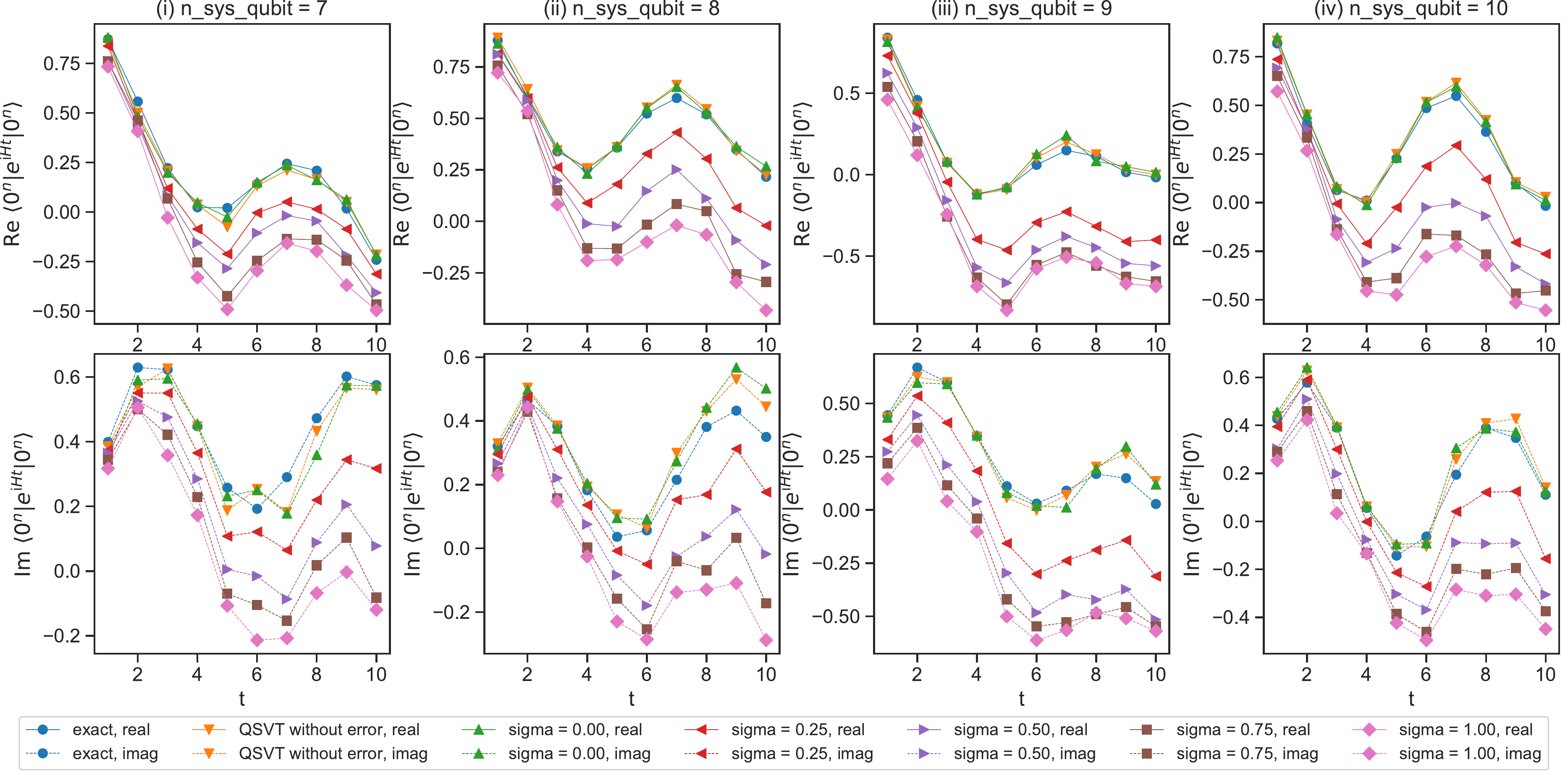}
    }
    \vspace{-10pt}
    \caption{Time series computed by using the \chbem model and QSVT. The details about phase factors are given in \cref{tab:param-ts} and the details about block-encoded matrices are given in \cref{fig:svd-matrix}. (a) Results obtained on IBM Q backends. The probability of CNOT is 0.1 when generating random circuits. (b) Results obtained on QVM with architecture and noise model retrieved from the IBM Q backend \textsf{ibmq\_16\_melbourne}. The probability of CNOT is set to 0.5.}
    \label{fig:ts}
\end{figure*}

\vspace{1em}
\noindent\textit{Time series analysis:} 
The numerical results regarding the time series analysis are shown in \cref{fig:ts}. The results in \cref{fig:ts-ibmq} are obtained on the IBM Q backends. When the number of system qubit is $1$ (the circuit uses 3 qubits in total), the features of the time series obtained from the quantum device can agree qualitatively with the exact solution. However, as the number of system qubits increases to $5$, the result is dominated by the quantum noise. In order to investigate the performance of larger systems and the effects of noise, we use the tunable QVM instead in \cref{fig:ts-qasm}.  The length of phase factors is chosen adaptively as $t$ increase in order to reduce the error of the polynomial approximation (details in \cref{tab:param-ts}). When the noise level $\sigma$ is tuned to $0$, the results from the QSVT circuit is uniformly accurate for all $t$ in $[1,10]$. However, since the circuit depth needs to increase proportionally to $t$, when the noise level is not $0$, the error is significantly larger as $t$ increases. We also remark that our custom noise model with $\sigma=1$ only contains partial kinds of noise derived from the real IBM Q device (details in \cref{app:model}). Therefore, in \cref{fig:ts}, the time series computed on the real device behaves much worse than that computed from the simulation using the noise model with $\sigma=1$ on the QVM.

\begin{figure}[htbp]
    \centering
    \includegraphics[width=8cm]{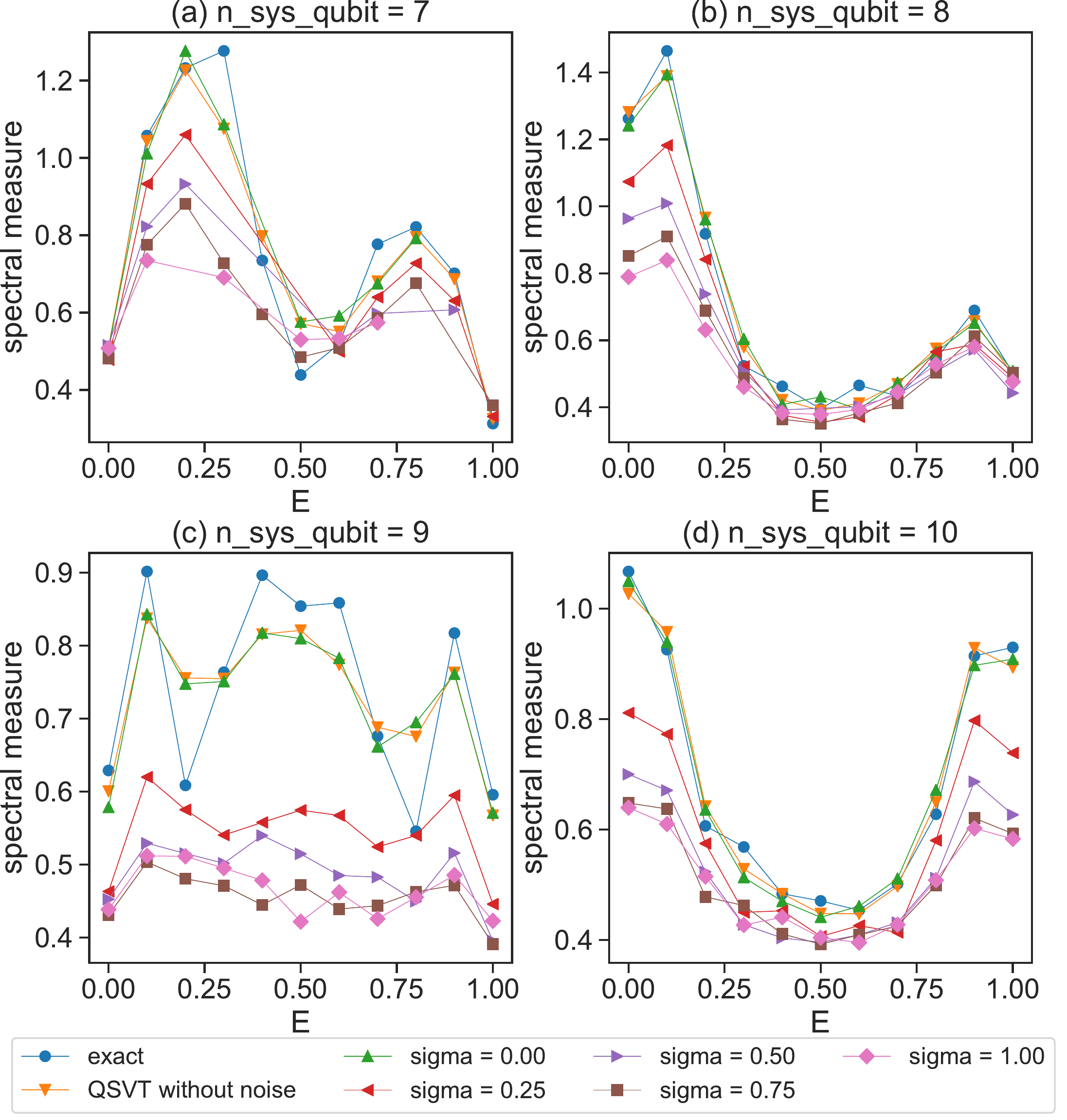}
    \caption{Spectral measure computed by using the \chbem model and QSVT. The numerical results are obtained by running on QVM with architecture and noise model retrieved from IBM Q backend \textsf{ibmq\_16\_melbourne}. For each number of system qubits, we draw a \chbem randomly according to the backend architecture. The probability of CNOT is 0.5 when generating random circuits. The details about phase factors are given in \cref{tab:param-sm} and the details about block-encoded matrices are given in \cref{fig:svd-matrix}.}
    \label{fig:sm-qasm}
\end{figure}

\vspace{1em}
\noindent\textit{Spectral measure:} In the limit when $\eta\to 0$, the spectral measure is given by the summation of Dirac-$\delta$ functions. Hence when $n$ is small, the spectral measure has sharp features even when $\eta$ takes a finite value, which in turn requires a polynomial of higher degree to resolve. So we only consider  spectral measures for $n=7,8,9,10$ on QVM, and the size of $\mathfrak{H}$ ranges from $128$ to $1024$. The numerical results of spectral measures are presented in \cref{fig:sm-qasm}. The spectral measures exhibit rather different features. This is not so much related to the number of system qubits, and is mostly due to the specific instance of the \hbem. In all cases, the quantum algorithm can capture at least the qualitative features of the spectral measures, though the noise plays an important role particularly for the instance $n=9$. We also remark that the functionality of the QSVT circuit depends only on the set of phase factors as $E$ sweeps across the spectrum.  The length of corresponding QSVT phase factors is set to $11$ for each point. In \cref{tab:param-sm} and \cref{tab:param-sm-large}, for points closer to the middle of the spectrum ($E=0.5$), we observe that a polynomial of larger degree is needed to achieve the same accuracy in approximation. Therefore, it can be seen from \cref{fig:sm-qasm} that the deviation between the value given by noiseless QSVT and the exact one increases as the parameter moves towards the middle of the spectrum. The choice of the circuit depth reflects the tradeoff between approximation error and the effects of noise. To illustrate this, we also compute the spectral measures in \cref{app:add-num} by using a deeper QSVT circuits. Although the deeper circuit can produce better results in the noiseless setting, the quality of the spectral measure can  significantly deteriorate in the presence of the quantum noise.

\begin{figure}[htbp]
    \centering
    \vspace{5pt}
    \subfigure[\label{fig:ta-ibmq}]{
        \includegraphics[width=.45\textwidth]{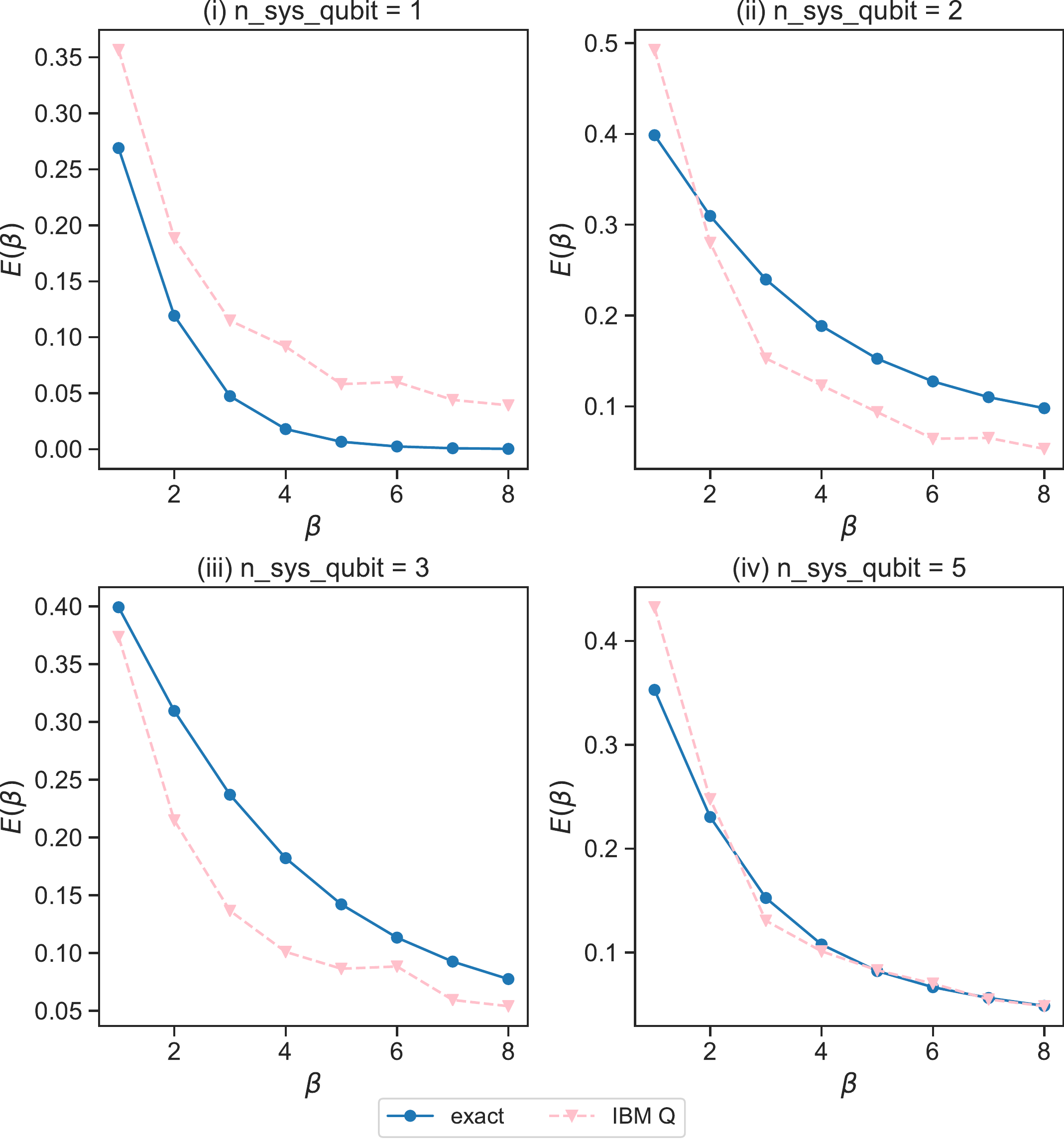}
    }
    \subfigure[\label{fig:ta-qasm}]{
        \includegraphics[width=.45\textwidth]{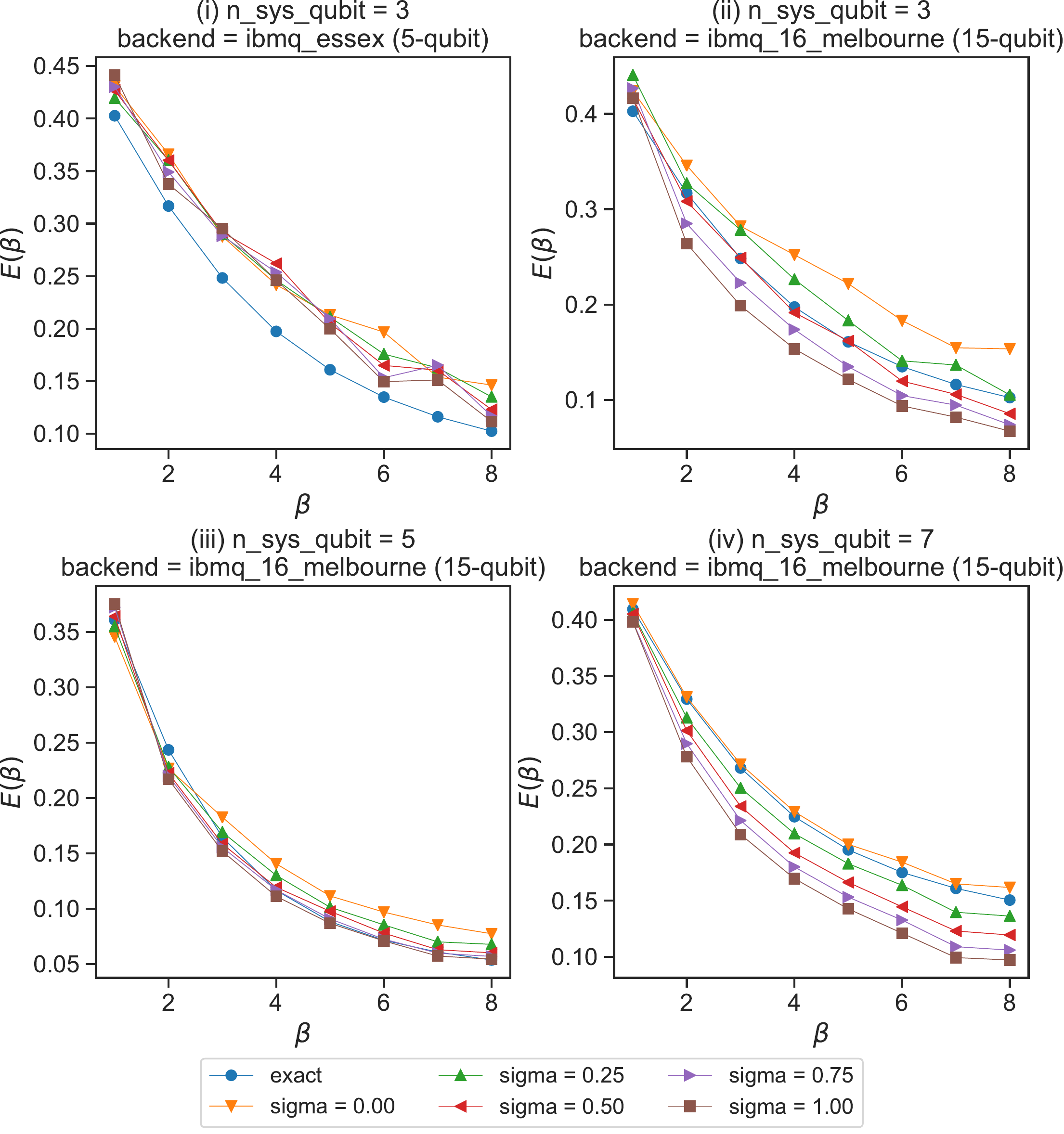}
    }
    \vspace{-10pt}
     \caption{Thermal average of the energy computed by using the \chbem model and QSVT. The details about phase factors are given in \cref{tab:param-ta} and the details about block-encoded matrices are given in \cref{fig:svd-matrix}. (a) Results obtained on IBM Q backends. The probability of CNOT is 0.1 when generating random circuits. (b) Results obtained on QVM. The probability of CNOT is set to 0.5.}
     \label{fig:ta}
 \end{figure}

\vspace{1em}
\noindent\textit{Thermal average of the energy:}
In \cref{fig:ta-ibmq}, we compute the thermal average of the energy of \hbem's on IBM Q backends for $n$ from $1$ to $5$. We reduce the length of QSVT phase factors as much as possible while balancing the approximation error and the quantum error, and the details can be found in \cref{tab:param-ta}. Since METTS is a Monte Carlo algorithm, we perform a sufficiently large number of steps to reduce the error due to the METTS algorithm, by monitoring the cumulative moving average of the thermal average of the energy in \cref{appfig:ta-cma} in \cref{app:add-num}.
Compared to the results above obtained on IBM Q, the results for the thermal average of the energy are somewhat surprisingly accurate in all cases. We also compute the thermal average of the energy on QVM to further investigate the effect of the noise for $n=3, 5$ and $7$. In \cref{fig:ta-qasm}(i) and (ii), the identical quantum computing task is emulated on different backends, and the only difference is due to the noise model. It is evident that the behaviour of the solution depends sensitive on the noise. On the other hand, we find that the solution remains remarkably accurate as $n$ increases, even when $\beta$ is relatively large. For most data points, the thermal average of the energy decreases monotonically with respect to $\beta$, and we observe that as $\sigma$ increases, the energy curve shifts downwards monotonically.

\section{Conclusion}
Analogous to the LINPACK benchmark for measuring the performance of classical supercomputers, we have proposed a quantum LINPACK benchmark to measure the performance of current and future quantum computers for scientific computing applications. The quantum LINPACK benchmark solves a well conditioned quantum linear system problem, which is enabled by the Hermitian RAndom Circuit Block-Encoded Matrix (\hbem) input model and the quantum singular value transformation (QSVT). The flexibility provided by the \hbem model also allows us to perform other linear algebra tasks already on near-term devices, such as computing spectral measures, and time series generated by a Hamiltonian simulation. We can also compute the thermal average of the energy, using a quantum version of the minimally entangled typical thermal state (METTS) algorithm. 

We perform these linear algebra tasks on IBM Q quantum devices, and quantum virtual machines with a tunable noise model. Although present quantum devices still suffer from high noise levels, it is nonetheless encouraging to observe that the solutions can already be qualitatively obtained in the noisy environment. Among all numerical tests, the thermal average of the energy computed via the METTS algorithm appears to be particularly robust with respect to noise. When designing the quantum circuit, we need to carefully choose the length of QSVT phase factors, by balancing the polynomial approximation error and the additional quantum error caused by the increase of the circuit depth.  Our numerical tests are currently limited to matrices up to $10$ qubits (the corresponding matrix size is $1024$) in order to compare with numerically exact results obtained on classical computers. However, the number of qubits can be directly increased to be much larger, especially on future devices with reduced noise level. 

Note that Google's quantum supremacy experiment is a benchmark problem (the linear cross-entropy test) and is also hard for any classical computer. The classical hardness is justified by that sampling a certain random quantum circuit (e.g. one that implements a chaotic quantum evolution) and generating heavy outputs is a difficult task on any classical computer \cite{EmersonWeinsteinSaracenoEtAl2003,HarrowLow2009,NahumRuhmanVijayEtAl2017,AaronsonArkhipov2011,BremnerMontanaroShepherd2016,Fujii2016,AaronsonChen2016,BoulandFeffermanNirkheEtAl2019,AaronsonGunn2019}. Although the quantum LINPACK benchmark also involves a supremacy-type random circuits, the success of the benchmark is defined by measuring only two ancilla qubits, and the task could therefore \emph{possibly} be ``classically spoofed'' when the block-encoding matrix $U_A$ is drawn from a known distribution, such as the Haar measure. We would like to therefore clarify that a quantum benchmark problem does not need to be classically hard in order to provide useful information on the performance of quantum computers. Indeed,  we expect that the success of quantum LINPACK benchmark is the \emph{minimal requirement} in order to achieve quantum advantages via linear algebra tasks such as QLSP. On the other hand, it is possible to go beyond the quantum LINPACK benchmark, and to formulate a more elaborate cross-entropy test related to QLSP that is also classically hard. This requires detailed studied of the statistical properties of truncated unitaries in the block-encoding model and QSVT problem, which will be our future work.

\section*{Acknowledgements}
This work was partially supported by a Google Quantum Research Award (Y.D.,L.L.), by the Department of Energy under Grant No. DE-AC02-05CH11231, No. DE-SC0017867, and the Quantum System Accelerator project (L.L.). 
This research used resources of the Oak Ridge Leadership Computing Facility, which is a DOE Office of Science User Facility supported under Contract No. DE-AC05-00OR22725.
We thank Yunchao Liu, Yu Tong and K. Birgitta Whaley for useful discussions.

\bibliographystyle{abbrvnat}

\begin{thebibliography}{46}
	\providecommand{\natexlab}[1]{#1}
	\providecommand{\url}[1]{\texttt{#1}}
	\expandafter\ifx\csname urlstyle\endcsname\relax
	\providecommand{\doi}[1]{doi: #1}\else
	\providecommand{\doi}{doi: \begingroup \urlstyle{rm}\Url}\fi
	
	\bibitem[IBM()]{IBMQ}
	URL \url{https://quantum-computing.ibm.com}.
	
	\bibitem[TOP()]{TOP500}
	URL \url{https://www.top500.org/project/}.
	
	\bibitem[Aaronson and Arkhipov(2011)]{AaronsonArkhipov2011}
	S.~Aaronson and A.~Arkhipov.
	\newblock The computational complexity of linear optics.
	\newblock In \emph{Proceedings of the forty-third annual ACM symposium on
		Theory of computing}, pages 333--342, 2011.
	
	\bibitem[Aaronson and Chen(2016)]{AaronsonChen2016}
	S.~Aaronson and L.~Chen.
	\newblock Complexity theoretic foundations of quantum supremacy experiments.
	\newblock \emph{arXiv:1612.05903}, 2016.
	
	\bibitem[Aaronson and Gunn(2019)]{AaronsonGunn2019}
	S.~Aaronson and S.~Gunn.
	\newblock {On the Classical Hardness of Spoofing Linear Cross-Entropy
		Benchmarking}.
	\newblock pages 1--7, 2019.
	
	\bibitem[Abraham et~al.(2019)]{Qiskit}
	H.~Abraham et~al.
	\newblock Qiskit: An open-source framework for quantum computing, 2019.
	
	\bibitem[An and Lin(2019)]{AnLin2019}
	D.~An and L.~Lin.
	\newblock Quantum linear system solver based on time-optimal adiabatic quantum
	computing and quantum approximate optimization algorithm.
	\newblock \emph{arXiv:1909.05500}, 2019.
	
	\bibitem[Arute et~al.(2019)Arute, Arya, Babbush, Bacon, Bardin, Barends,
	Biswas, Boixo, Brandao, Buell, et~al.]{AruteAryaBabbushEtAl2019}
	F.~Arute, K.~Arya, R.~Babbush, D.~Bacon, J.~C. Bardin, R.~Barends, R.~Biswas,
	S.~Boixo, F.~G. Brandao, D.~A. Buell, et~al.
	\newblock Quantum supremacy using a programmable superconducting processor.
	\newblock \emph{Nature}, 574\penalty0 (7779):\penalty0 505--510, 2019.
	
	\bibitem[Barenco et~al.(1995)Barenco, Bennett, Cleve, DiVincenzo, Margolus,
	Shor, Sleator, Smolin, and Weinfurter]{BarencoBennettCleveEtAl1995}
	A.~Barenco, C.~H. Bennett, R.~Cleve, D.~P. DiVincenzo, N.~Margolus, P.~Shor,
	T.~Sleator, J.~A. Smolin, and H.~Weinfurter.
	\newblock Elementary gates for quantum computation.
	\newblock \emph{Phys. Rev. A}, 52\penalty0 (5):\penalty0 3457, 1995.
	
	\bibitem[Berry et~al.(2015)Berry, Childs, and Kothari]{BerryChildsKothari2015}
	D.~W. Berry, A.~M. Childs, and R.~Kothari.
	\newblock Hamiltonian simulation with nearly optimal dependence on all
	parameters.
	\newblock \emph{Proceedings of the 56th IEEE Symposium on Foundations of
		Computer Science}, pages 792--809, 2015.
	
	\bibitem[Blume-Kohout et~al.(2017)Blume-Kohout, Gamble, Nielsen, Rudinger,
	Mizrahi, Fortier, and Maunz]{Blume-KohoutGambleNielsenEtAl2017}
	R.~Blume-Kohout, J.~K. Gamble, E.~Nielsen, K.~Rudinger, J.~Mizrahi, K.~Fortier,
	and P.~Maunz.
	\newblock Demonstration of qubit operations below a rigorous fault tolerance
	threshold with gate set tomography.
	\newblock \emph{Nat. Commun.}, 8\penalty0 (1):\penalty0 1--13, 2017.
	
	\bibitem[Boixo et~al.(2018)Boixo, Isakov, Smelyanskiy, Babbush, Ding, Jiang,
	Bremner, Martinis, and Neven]{BoixoIsakovSmelyanskiyEtAl2018}
	S.~Boixo, S.~V. Isakov, V.~N. Smelyanskiy, R.~Babbush, N.~Ding, Z.~Jiang, M.~J.
	Bremner, J.~M. Martinis, and H.~Neven.
	\newblock Characterizing quantum supremacy in near-term devices.
	\newblock \emph{Nat. Phys.}, 14\penalty0 (6):\penalty0 595--600, 2018.
	
	\bibitem[Bouland et~al.(2019)Bouland, Fefferman, Nirkhe, and
	Vazirani]{BoulandFeffermanNirkheEtAl2019}
	A.~Bouland, B.~Fefferman, C.~Nirkhe, and U.~Vazirani.
	\newblock On the complexity and verification of quantum random circuit
	sampling.
	\newblock \emph{Nat. Phys.}, 15\penalty0 (2):\penalty0 159--163, 2019.
	
	\bibitem[Bravo-Prieto et~al.(2019)Bravo-Prieto, LaRose, Cerezo, Subasi, Cincio,
	and Coles]{Bravo-PrietoLaRoseCerezoEtAl2019}
	C.~Bravo-Prieto, R.~LaRose, M.~Cerezo, Y.~Subasi, L.~Cincio, and P.~J. Coles.
	\newblock Variational quantum linear solver: A hybrid algorithm for linear
	systems.
	\newblock \emph{arXiv:1909.05820}, 2019.
	
	\bibitem[Bremner et~al.(2016)Bremner, Montanaro, and
	Shepherd]{BremnerMontanaroShepherd2016}
	M.~J. Bremner, A.~Montanaro, and D.~J. Shepherd.
	\newblock Average-case complexity versus approximate simulation of commuting
	quantum computations.
	\newblock \emph{Phys. Rev. Lett.}, 117\penalty0 (8):\penalty0 080501, 2016.
	
	\bibitem[Cao et~al.(2013)Cao, Papageorgiou, Petras, Traub, and
	Kais]{CaoPapageorgiouPetrasEtAl2013}
	Y.~Cao, A.~Papageorgiou, I.~Petras, J.~Traub, and S.~Kais.
	\newblock Quantum algorithm and circuit design solving the poisson equation.
	\newblock \emph{New J. Phys.}, 15\penalty0 (1):\penalty0 013021, 2013.
	
	\bibitem[Casares and Martin-Delgado(2019)]{CasaresMartin-Delgado2019}
	P.~Casares and M.~Martin-Delgado.
	\newblock A quantum ip predictor-corrector algorithm for linear programming.
	\newblock \emph{arXiv preprint arXiv:1902.06749}, 2019.
	
	\bibitem[Chakraborty et~al.(2018)Chakraborty, Gily{\'e}n, and
	Jeffery]{ChakrabortyGilyenJeffery2018}
	S.~Chakraborty, A.~Gily{\'e}n, and S.~Jeffery.
	\newblock The power of block-encoded matrix powers: improved regression
	techniques via faster hamiltonian simulation.
	\newblock \emph{arXiv:1804.01973}, 2018.
	
	\bibitem[Childs et~al.(2017)Childs, Kothari, and Somma]{ChildsKothariSomma2017}
	A.~M. Childs, R.~Kothari, and R.~D. Somma.
	\newblock Quantum algorithm for systems of linear equations with exponentially
	improved dependence on precision.
	\newblock \emph{SIAM J. Comput.}, 46:\penalty0 1920--1950, 2017.
	
	\bibitem[Cross et~al.(2019)Cross, Bishop, Sheldon, Nation, and
	Gambetta]{CrossBishopSheldonEtAl2019}
	A.~W. Cross, L.~S. Bishop, S.~Sheldon, P.~D. Nation, and J.~M. Gambetta.
	\newblock Validating quantum computers using randomized model circuits.
	\newblock \emph{Physical Review A}, 100\penalty0 (3):\penalty0 032328, 2019.
	
	\bibitem[Dong et~al.(2020)Dong, Meng, Whaley, and Lin]{DongMengWhaleyEtAl}
	Y.~Dong, X.~Meng, K.~B. Whaley, and L.~Lin.
	\newblock Efficient phase factor evaluation in quantum signal processing.
	\newblock \emph{arXiv:2002.11649}, 2020.
	
	\bibitem[Dongarra et~al.(2003)Dongarra, Luszczek, and
	Petitet]{DongarraLuszczekPetitet2003}
	J.~J. Dongarra, P.~Luszczek, and A.~Petitet.
	\newblock {The LINPACK benchmark: past, present and future}.
	\newblock \emph{Concurrency and Computation: practice and experience},
	15\penalty0 (9):\penalty0 803--820, 2003.
	
	\bibitem[Emerson et~al.(2003)Emerson, Weinstein, Saraceno, Lloyd, and
	Cory]{EmersonWeinsteinSaracenoEtAl2003}
	J.~Emerson, Y.~S. Weinstein, M.~Saraceno, S.~Lloyd, and D.~G. Cory.
	\newblock Pseudo-random unitary operators for quantum information processing.
	\newblock \emph{Science}, 302\penalty0 (5653):\penalty0 2098--2100, 2003.
	
	\bibitem[Erhard et~al.(2019)Erhard, Wallman, Postler, Meth, Stricker, Martinez,
	Schindler, Monz, Emerson, and Blatt]{ErhardWallmanPostlerEtAl2019}
	A.~Erhard, J.~J. Wallman, L.~Postler, M.~Meth, R.~Stricker, E.~A. Martinez,
	P.~Schindler, T.~Monz, J.~Emerson, and R.~Blatt.
	\newblock {Characterizing large-scale quantum computers via cycle
		benchmarking}.
	\newblock \emph{Nat. Commun.}, 10\penalty0 (1):\penalty0 5347, 2019.
	
	\bibitem[Fujii(2016)]{Fujii2016}
	K.~Fujii.
	\newblock Noise threshold of quantum supremacy.
	\newblock \emph{arXiv:1610.03632}, 2016.
	
	\bibitem[Gily{\'e}n et~al.(2018)Gily{\'e}n, Su, Low, and
	Wiebe]{GilyenSuLowEtAl2018}
	A.~Gily{\'e}n, Y.~Su, G.~H. Low, and N.~Wiebe.
	\newblock Quantum singular value transformation and beyond: exponential
	improvements for quantum matrix arithmetics.
	\newblock \emph{arXiv:1806.01838}, 2018.
	
	\bibitem[Gily{\'e}n et~al.(2019)Gily{\'e}n, Su, Low, and
	Wiebe]{GilyenSuLowEtAl2019}
	A.~Gily{\'e}n, Y.~Su, G.~H. Low, and N.~Wiebe.
	\newblock Quantum singular value transformation and beyond: exponential
	improvements for quantum matrix arithmetics.
	\newblock In \emph{Proceedings of the 51st Annual ACM SIGACT Symposium on
		Theory of Computing}, pages 193--204, 2019.
	
	\bibitem[Giovannetti et~al.(2008)Giovannetti, Lloyd, and
	Maccone]{GiovannettiLloydMaccone2008}
	V.~Giovannetti, S.~Lloyd, and L.~Maccone.
	\newblock Quantum random access memory.
	\newblock \emph{Phys. Rev. Lett.}, 100\penalty0 (16):\penalty0 160501, 2008.
	
	\bibitem[Haah(2019)]{Haah2019}
	J.~Haah.
	\newblock Product decomposition of periodic functions in quantum signal
	processing.
	\newblock \emph{Quantum}, 3:\penalty0 190, 2019.
	
	\bibitem[Harrow and Low(2009)]{HarrowLow2009}
	A.~W. Harrow and R.~A. Low.
	\newblock Random quantum circuits are approximate 2-designs.
	\newblock \emph{Commun. Math. Phys.}, 291\penalty0 (1):\penalty0 257--302,
	2009.
	
	\bibitem[Harrow et~al.(2009)Harrow, Hassidim, and
	Lloyd]{HarrowHassidimLloyd2009}
	A.~W. Harrow, A.~Hassidim, and S.~Lloyd.
	\newblock Quantum algorithm for linear systems of equations.
	\newblock \emph{Phys. Rev. Lett.}, 103:\penalty0 150502, 2009.
	
	\bibitem[Lin and Tong(2019)]{LinTong2019}
	L.~Lin and Y.~Tong.
	\newblock Solving quantum linear system problem with near-optimal complexity.
	\newblock \emph{arXiv:1910.14596}, 2019.
	
	\bibitem[Magesan et~al.(2011)Magesan, Gambetta, and
	Emerson]{MagesanGambettaEmerson2011}
	E.~Magesan, J.~M. Gambetta, and J.~Emerson.
	\newblock Scalable and robust randomized benchmarking of quantum processes.
	\newblock \emph{Phys. Rev. Lett.}, 106\penalty0 (18):\penalty0 180504, 2011.
	
	\bibitem[McKay et~al.(2017)McKay, Wood, Sheldon, Chow, and
	Gambetta]{MckayWoodChrisEtAl2017}
	D.~C. McKay, C.~J. Wood, S.~Sheldon, J.~M. Chow, and J.~M. Gambetta.
	\newblock Efficient z gates for quantum computing.
	\newblock \emph{Phys. Rev. A}, 96\penalty0 (2):\penalty0 022330, 2017.
	
	\bibitem[Motta et~al.(2020)Motta, Sun, Tan, O'Rourke, Ye, Minnich, Brand{\~a}o,
	and Chan]{MottaSunTanEtAl2020}
	M.~Motta, C.~Sun, A.~T. Tan, M.~J. O'Rourke, E.~Ye, A.~J. Minnich, F.~G.
	Brand{\~a}o, and G.~K.-L. Chan.
	\newblock Determining eigenstates and thermal states on a quantum computer
	using quantum imaginary time evolution.
	\newblock \emph{Nat. Phys.}, 16\penalty0 (2):\penalty0 205--210, 2020.
	
	\bibitem[Nahum et~al.(2017)Nahum, Ruhman, Vijay, and
	Haah]{NahumRuhmanVijayEtAl2017}
	A.~Nahum, J.~Ruhman, S.~Vijay, and J.~Haah.
	\newblock Quantum entanglement growth under random unitary dynamics.
	\newblock \emph{Phys. Rev. X}, 7\penalty0 (3):\penalty0 031016, 2017.
	
	\bibitem[Proctor et~al.(2020)Proctor, Rudinger, Young, Nielsen, and
	Blume-Kohout]{ProctorRudingerYoungEtAl2020}
	T.~Proctor, K.~Rudinger, K.~Young, E.~Nielsen, and R.~Blume-Kohout.
	\newblock Measuring the capabilities of quantum computers.
	\newblock 2020.
	
	\bibitem[Saeedi and Pedram(2013)]{SaeediPedram2013}
	M.~Saeedi and M.~Pedram.
	\newblock Linear-depth quantum circuits for n-qubit toffoli gates with no
	ancilla.
	\newblock \emph{Phys. Rev. A}, 87\penalty0 (6):\penalty0 062318, 2013.
	
	\bibitem[Somma(2019)]{Somma2019}
	R.~D. Somma.
	\newblock Quantum eigenvalue estimation via time series analysis.
	\newblock \emph{New J. Phys.}, 21\penalty0 (12):\penalty0 123025, 2019.
	
	\bibitem[Stoudenmire and White(2010)]{StoudenmireWhite2010}
	E.~Stoudenmire and S.~R. White.
	\newblock Minimally entangled typical thermal state algorithms.
	\newblock \emph{New J. Phys.}, 12\penalty0 (5):\penalty0 055026, 2010.
	
	\bibitem[Suba{\c{s}}{\i} et~al.(2019)Suba{\c{s}}{\i}, Somma, and
	Orsucci]{SubasiSommaOrsucci2019}
	Y.~Suba{\c{s}}{\i}, R.~D. Somma, and D.~Orsucci.
	\newblock Quantum algorithms for systems of linear equations inspired by
	adiabatic quantum computing.
	\newblock \emph{Phys. Rev. Lett.}, 122:\penalty0 060504, 2019.
	
	\bibitem[Tang(2019)]{Tang2019}
	E.~Tang.
	\newblock A quantum-inspired classical algorithm for recommendation systems.
	\newblock In \emph{Proceedings of the 51st Annual ACM SIGACT Symposium on
		Theory of Computing}, pages 217--228, 2019.
	
	\bibitem[Tong et~al.(2020)Tong, An, Wiebe, and Lin]{TongAnWiebeEtAl2020}
	Y.~Tong, D.~An, N.~Wiebe, and L.~Lin.
	\newblock Fast inversion, preconditioned quantum linear system solvers, and
	fast evaluation of matrix functions.
	\newblock \emph{arXiv:2008.13295}, 2020.
	
	\bibitem[White(2009)]{White2009}
	S.~R. White.
	\newblock Minimally entangled typical quantum states at finite temperature.
	\newblock \emph{Phys. Rev. Lett.}, 102\penalty0 (19):\penalty0 190601, 2009.
	
	\bibitem[Wossnig et~al.(2018)Wossnig, Zhao, and
	Prakash]{WossnigZhaoPrakash2018}
	L.~Wossnig, Z.~Zhao, and A.~Prakash.
	\newblock Quantum linear system algorithm for dense matrices.
	\newblock \emph{Phys. Rev. Lett.}, 120\penalty0 (5):\penalty0 050502, 2018.
	
	\bibitem[Xu et~al.(2019)Xu, Sun, Endo, Li, Benjamin, and
	Yuan]{XuSunEndoEtAl2019}
	X.~Xu, J.~Sun, S.~Endo, Y.~Li, S.~C. Benjamin, and X.~Yuan.
	\newblock Variational algorithms for linear algebra.
	\newblock \emph{arXiv:1909.03898}, 2019.
	
\end{thebibliography}

\widetext
\clearpage
\appendix

\setcounter{equation}{0}
\setcounter{figure}{0}
\setcounter{table}{0}

\renewcommand{\theequation}{A\arabic{equation}}
\renewcommand{\thefigure}{A\arabic{figure}}
\renewcommand{\thetable}{A\arabic{table}}

\section{Derivation of \hbem}\label{sec:hracbem}

To derive \cref{eqn:hracbem}, we start from a state $\ket{0}\ket{0}\ket{\psi}$, and follow the circuit in \cref{fig:qsp_circuit_hermitian_racbem}. After applying $U_A$, the state becomes
\begin{equation}
\frac{1}{\sqrt{2}} \left(e^{\I \varphi_0}\ket{0}+e^{-\I \varphi_0}\ket{1}\right)(\ket{0}A\ket{\psi}+\ket{1}\ket{\perp}).
\label{eqn:hracbem_tmp1}
\end{equation}
Here $\ket{\perp}$ is an $n$-qubit state defined through the relation
\[
U_A \ket{0}\ket{\psi}=\ket{0}A\ket{\psi} + \ket{1}\ket{\perp}.
\]
Therefore
\[
\ket{0}\ket{\psi}=U_A^{\dag}U_A \ket{0}\ket{\psi}=\ket{0} (A^{\dag}A)\ket{\psi}+\ket{1}\ket{\perp'}+U_A^{\dag} \ket{1}\ket{\perp},
\]
or
\begin{equation}
U_A^{\dag} \ket{1}\ket{\perp}=\ket{0}(I_n-A^{\dag}A)\ket{\psi}-\ket{1}\ket{\perp'}.
\label{eqn:hracbem_tmp2}
\end{equation}
Here $\ket{\perp'}$ is another $n$-qubit state. Via the relation \eqref{eqn:hracbem_tmp2}, after applying $U_A^{\dag}$, the state \eqref{eqn:hracbem_tmp1} is transformed to
\begin{equation}
\begin{aligned}
&\frac{1}{\sqrt{2}} \left(e^{\I (\varphi_0+\varphi_1)}\ket{0}+e^{-\I (\varphi_0+\varphi_1)}\ket{1}\right)(\ket{0}A^{\dag}A\ket{\psi}+\ket{1}\ket{\perp'})\\
+&\frac{1}{\sqrt{2}} \left(e^{\I (\varphi_0-\varphi_1)}\ket{0}+e^{-\I (\varphi_0-\varphi_1)}\ket{1}\right)(\ket{0}(I_n-A^{\dag}A)\ket{\psi}-\ket{1}\ket{\perp'}).
\end{aligned}
\label{eqn:hracbem_tmp3}
\end{equation}
Finally, carrying out the remaining operations of the circuit in \cref{fig:qsp_circuit_hermitian_racbem}, and applying $\bra{0^2}\otimes I_n$, we obtain the form in  \cref{eqn:hracbem}.

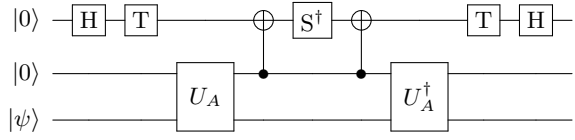
\begin{figure}[htbp]
\begin{center}
\[
\Qcircuit @C=0.8em @R=1.em {
 \lstick{\ket{0}}& \gate{\mathrm{H}}  & \gate{\mathrm{T}} & \qw&\targ & \gate{\mathrm{S}^{\dag}} & \targ & \qw&\gate{\mathrm{T}} & \gate{\mathrm{H}}&\qw  \\
\lstick{\ket{0}}& \qw  & \qw & \multigate{1}{U_A} & \ctrl{-1} & \qw & \ctrl{-1} & \multigate{1}{U^{\dag}_A}    &\qw& \qw&\qw \\
\lstick{\ket{\psi}}& \qw &\qw  &\ghost{U_A} &\qw&\qw&\qw&\ghost{U^{\dag}_A}&\qw&\qw&\qw
}
\]
\end{center}
\caption{Quantum circuit for generating a $(1,2,0)$-block-encoding of a \chbem from a $(1,1,0)$-block-encoding $U_A$ and its Hermitian conjugate. $\mathrm{H}$ is the Hadamard gate, and $\mathrm{Z}$ is the Pauli-Z gate.}
\label{fig:qsp_circuit_canonical_hermitian_racbem}
\end{figure}

 The quantum circuit for representing a \chbem can be simplified using \cref{fig:qsp_circuit_canonical_hermitian_racbem}, still denoted by $U_{\mathfrak{H}}$. Here the two phase shift gates are
\[
\mathrm{S}=\begin{pmatrix}
1 & 0\\
0 & \I
\end{pmatrix}, \quad
\mathrm{T}=\begin{pmatrix}
1 & 0\\
0 & e^{\I \frac{\pi}{4}}
\end{pmatrix}.
\]
Furthermore, we removed the controlled-NOT gates near the Hadamard gates, and replaced the controlled-NOT gates by the standard 
CNOT gate controlled on $1$. 

Following the same line of calculation above, starting from $\ket{0}\ket{0}\ket{\psi}$, the state is transformed to
\[
\frac{1}{\sqrt{2}} \left(\ket{0}+e^{\I \frac{\pi}{4}}\ket{1}\right)(\ket{0}A\ket{\psi}+\ket{1}\ket{\perp})
\]
after applying $U_A$, and then to
\[
\begin{aligned}
&\frac{1}{\sqrt{2}} \left(\ket{0}+e^{-\I \frac{\pi}{4}}\ket{1}\right)(\ket{0}A^{\dag}A\ket{\psi}+\ket{1}\ket{\perp'})\\
+&\frac{1}{\sqrt{2}} \left(e^{-\I \frac{\pi}{2}}\ket{0}+e^{\I \frac{\pi}{4}}\ket{1}\right)(\ket{0}(I_n-A^{\dag}A)\ket{\psi}-\ket{1}\ket{\perp'}),
\end{aligned}
\]
after applying $U_A^{\dag}$. Finally, applying the remaining $\mathrm{T}$ and $\mathrm{H}$ gates, as well as $\bra{0^2}\otimes I_n$, we obtain the form
\[
\mathfrak{H}\ket{\psi}=(\bra{0^2}\otimes I_n)U_{\mathfrak{H}}\ket{0^2}\ket{\psi}=A^{\dag}A\ket{\psi}.
\]

\section{Optimization based method for finding phase factors}\label{sec:optimization}
In a recent work \cite{DongMengWhaleyEtAl}, the authors proposed an optimization-based algorithm for finding phase factors. This is an optimization problem involving matrices only in $\text{SU}(2)$, and is independent of the number of system qubits $n$. The goal is to find a parameterized matrix 
$U_\Phi(x) = e^{\I \phi_0 \mathrm{Z}} \prod_{j=1}^\qspdeg \left[ e^{\I \arccos(x) \mathrm{X}} e^{\I \phi_j \mathrm{Z}} \right]$, where phase factors $\Phi := (\phi_0, \cdots, \phi_d)$ are related to those defined in the main text by the following relation:
\begin{equation}
\varphi_i=\begin{cases}
\phi_0+\frac{\pi}{4}, & i=0,\\
\phi_i+\frac{\pi}{2}, & 1\le i\le \qspdeg-1,\\
\phi_n+\frac{\pi}{4}, & i=\qspdeg.\\
\end{cases}
\label{eqn:phi_varphi_relation}
\end{equation}
The objective function of the optimization problem is the distance between a real polynomial $f$ and $\Re [\langle0|U_\Phi(\cdot)|0\rangle]$. The polynomial $f$ is of degree $\qspdeg$ and its parity is $(\qspdeg \mod 2)$.  Hence the polynomial can be determined via  $\tilde \qspdeg = \lceil \frac{\qspdeg+1}{2} \rceil$ degrees of freedom. Then according to \cite{DongMengWhaleyEtAl}, the objective function can be defined as
\begin{equation}
    L(\Phi) = \frac{1}{\tilde \qspdeg} \sum_{j=1}^{\tilde \qspdeg}\left|\Re \left[\left\langle 0\left|U_{\Phi}\left(x_{j}\right)\right| 0\right\rangle\right]-f\left(x_{j}\right)\right|^{2}
\end{equation}
where $x_j=\cos\left(\frac{(2j-1)\pi}{4\tilde \qspdeg}\right),\,j=1,\dots,{\tilde \qspdeg}$ are the positive roots of the Chebyshev polynomial $T_{2\tilde \qspdeg}(x)$.  We may further restrict the set of phase factors to be centrally symmetric. The optimization can be implemented using the standard L-BFGS algorithm, and the running time of the algorithm scales only quadratically with respect to $d$. 

To approximate a generic real smooth function of definite parity, we can use either orthogonal projection onto the set of Chebyshev polynomials on $[-1,1]$, or use the Remez algorithm to compute the best polynomial approximation on the given subinterval of $[-1,1]$. This streamlines the procedure of finding the phase factors, and such a procedure can be used to identify phase factors for a very large polynomial degree ($d>10000$) with high precision.

\section{Simulation models}\label{app:model}
\vspace{1em}
\noindent\textit{Architecture of quantum devices:} The architecture of near-term quantum devices is characterized by several features including the coupling map, basic gates and noise error rates. The coupling map is given by a directed graph $G=\langle V,E\rangle$, and the coupling maps of quantum devices provided by IBM Q are always symmetric. The nodes (vertices) $V$ are the set of available qubits on the quantum device, and the edges $E$ specify the set of CNOT gates between qubit pairs that can be directly implemented on the device. The two nodes associated with an edge are assigned to be the control qubit and the target qubit of the CNOT gate, respectively. Basic gates are the building units of quantum circuits that can be directly implemented on the quantum device. If a quantum circuit involves more complicated quantum gates, these gates must first be decomposed into the composition of basic gates by a quantum transpiler before implementation on the quantum hardware. The noise error rate is a measure of the strength of noise on basic gates acting on permitted each qubit or qubit pair. 

On quantum computing backends provided by IBM Q, the basic gates are identity gate, CNOT gate, U1, U2 and U3 gate. Up to a global phase factor, U3 gate is
\[
\mathrm{U}_3(\theta,\phi,\lambda) = R_z(\phi+3\pi)R_x(\pi/2)R_z(\theta+\pi)R_x(\pi/2)R_z(\lambda),
\]
which is a generic single-qubit operation parameterized by three Euler angles. The U1 and U2 gates are defined by restricting to one or two Z-rotation angles respectively, i.e. 
\[
\mathrm{U}_1(\lambda) = R_z(\lambda), \quad \mathrm{U}_2(\phi, \lambda) = R_z(\phi + \pi/2) R_x(\pi/2) R_z(\lambda-\pi/2).
\]
The U3 gate can be used to generate arbitrary single-qubit operation \cite{MckayWoodChrisEtAl2017}. Moreover, in the absence of error correction, the Z-rotation gates can be implemented virtually with in principle negligible error rate. Hence, the error rate of U3 gate is mainly contributed by two X-rotations. Although the U1 and U2 gate are specific cases of the U3 gate, they are provided individually for the consideration of reducing error rate, since they involve only zero or one X-rotation operation, respectively. Therefore we exclude U3 from the basic gate set and draw random circuits accordingly with respect to the coupling map and our restricted basic gates. Note that phase gate S and $\pi/8$-gate T can be implemented as U1 gates, and Hadamard gate H is a U2 gate. Our restricted basic gate set still has universal representability. The control gates involved in the QSVT circuit can also be directly implemented using the restricted basic gate set, and hence our implementation reduces the usage of noisy U3 gates as much as possible.

\vspace{1em}
\noindent\textit{Custom random circuit generator:}
The \textsf{Qiskit} package provides a utility routine to generate random circuits. However, it does not take into account the coupling map and basic gates available to the target backend, which can be highly inefficient and error prone. In particular, a CNOT gate cannot be directly implemented unless the two qubits are connected according to the coupling map. Therefore the random circuit generated completely randomly cannot be executed directly on the quantum hardware before invoking a quantum transpiler. As a result, we designed a custom random circuit generator in \cref{alg:custom-rqc}. The resulting random circuit can be directly implemented on the quantum hardware without the need of using a quantum transpiler. Note that if two U1 gates appear consecutively, they can be combined together if needed.

\begin{figure}
    \centering 
        \begin{center}
            \scalebox{0.575}{$A = \left(\begin{array}{*{8}{r}}
0.096+0.256\I & -0.041+0.058\I & 0.096-0.224\I & 0.120+0.061\I & -0.138-0.054\I & 0.013+0.052\I & 0.189-0.099\I & 0.152-0.166\I \\ 
0.143-0.023\I & 0.001-0.335\I & 0.046-0.237\I & -0.056+0.007\I & 0.063+0.016\I & 0.079-0.063\I & 0.017+0.276\I & -0.046+0.007\I \\ 
0.054-0.017\I & -0.073+0.149\I & -0.002-0.063\I & -0.128+0.128\I & -0.371+0.048\I & -0.163-0.102\I & -0.069-0.069\I & 0.126+0.037\I \\ 
-0.043-0.208\I & -0.156-0.170\I & 0.189-0.080\I & -0.090+0.142\I & -0.057+0.075\I & 0.252+0.080\I & 0.150+0.057\I & 0.098-0.043\I \\ 
0.145+0.178\I & -0.325+0.125\I & 0.114+0.242\I & -0.136-0.316\I & 0.145+0.255\I & -0.120-0.335\I & -0.046+0.295\I & -0.142-0.184\I \\ 
-0.117+0.149\I & -0.101+0.338\I & -0.213-0.018\I & -0.474+0.081\I & -0.036-0.121\I & 0.444+0.147\I & -0.198+0.035\I & -0.091-0.054\I \\ 
-0.063+0.305\I & 0.001-0.145\I & -0.177+0.045\I & -0.209-0.150\I & -0.041+0.296\I & 0.046+0.082\I & 0.387-0.051\I & -0.430+0.233\I \\ 
-0.093-0.127\I & 0.254+0.307\I & -0.144-0.265\I & -0.048-0.353\I & 0.023+0.060\I & 0.085-0.156\I & 0.011+0.225\I & 0.249+0.420\I \\ 
\end{array}\right)$}
        \end{center}
        \begin{center}
            $\Sigma = \left(\begin{array}{cccccccc}
0.96439 & 0.88817 & 0.86418 & 0.74016 & 0.67243 & 0.50319 & 0.45951 & 0.26449
\end{array}\right)$
        \end{center}
    \caption{The relevant information of \bem circuit in \cref{fig:racbem_3qubit}. The upper matrix $A$ is the 3-qubit matrix block-encoded as the upper-left block, namely, identifying $q_0$ as the block-encoding qubit. The elements in the lower array are the singular values of the block-encoded matrix $A$.}
    \label{fig:my_label}
\end{figure}

\begin{algorithm}[htbp]
\caption{Random quantum circuit generation for a given quantum device}
\label{alg:custom-rqc}
\begin{algorithmic}[0]
\STATE \textbf{Input:} Coupling map $G = \langle V, E\rangle$ where $V$ is the set of qubits, $E$ is the set of qubit pairs on which CNOT is available, basic gates $\Gamma \subseteq \{ \mathrm{U1}, \mathrm{U2}, \mathrm{U3}, \mathrm{CNOT} \}$, the probability of choosing CNOT gate $p \in (0, 1)$, the circuit depth $\ell$.
\vspace{1em}
\STATE Set $t = 0$, initialize an empty quantum circuit $\mc{C}$.
\WHILE{$t < \ell$}
\STATE Reset $\langle V, E\rangle:=G$.
\WHILE{$V$ is not empty}
\STATE Draw a random number $r$ uniformly on $[0, 1]$.
\IF{$r \le p$ and $E$ is not empty}
\STATE $\mathrm{P} = \mathrm{CNOT}$. 
\STATE Pick a pair of qubits uniformly at random from $E$ as operands.
\ELSE
\STATE Pick an operation $\mathrm{P}$ uniformly at random from $\Gamma \backslash \{\mathrm{CNOT}\}$.
\STATE Pick $n_p$ angles uniformly at random on $[0, 2\pi)$ to form $\mathrm{P}$. $n_p = 1,2,3$ for U1, U2, U3 respectively.
\STATE Pick a qubit uniformly at random from $V$ as the operand.
\ENDIF
\STATE Apply $\mathrm{P}$ to its operand(s) in the circuit $\mc{C}$.

\STATE Remove the operand(s) from $V$; remove the qubit pairs in which the operand(s) involves from $E$.
\ENDWHILE
\STATE Set $t = t + 1$.
\ENDWHILE
\STATE \textbf{Return:} A random quantum circuit $\mc{C}$ whose depth is $\ell$ and each layer is fully filled by one/two-qubit gates.
\end{algorithmic}
\end{algorithm}

\vspace{1em}
\noindent\textit{Construction of the error model:} In order to elucidate the impact of noise magnitude, we construct an adjustable noise model as follows. We first retrieve noise models from IBM Q backends which are calibrated by the provider. The retrieved noise model is a \textsf{python} dictionary consisting of the information of each error type. There are some error modes which are quantum errors and can be characterized by a discrete probability distribution, for example, the bit flip error and the phase flip error. Meanwhile, readout errors in the measurement are also associated with a discrete probability distribution. We introduce a parameter $\sigma \in [0,1]$ to smoothly adjust the magnitude of such errors. The correct operation corresponds to the entry with the largest magnitude in the probability distribution. Then, all other entries are scaled by a factor $\sigma$, and the correct entry is adjusted accordingly to satisfy the normalization condition of the distribution. For instance, suppose a quantum error mode is given by the distribution vector $p = [0.90, 0.06, 0.04]$. Then, we identify the quantum operation associated with first entry as the correct operation. The scaled distribution vector is then $p_\sigma = [1-(1-0.90)\sigma, 0.06\sigma, 0.04\sigma]$. Therefore, when $\sigma = 1$, the scaled error mode is identical to that retrieved from the backends. When $\sigma = 0$, only the correct operation is applied with probability $1$, and the corresponding error symptoms are eliminated. 

However, in \textsf{Qiskit} there is another type of quantum error modes, referred to the ``Kraus error''. These error modes are not characterized by a discrete probability distribution, and the corresponding Kraus operator is always applied with probability $1$. Hence such ``Kraus errors'' cannot be adjusted using the same method illustrated above. For simplicity of implementation, we discarded such error modes in our noise dictionary. Hence the noise level of our model can be lower than that retrieved directly from the quantum device, even when $\sigma=1$.  Nonetheless, our numerical results demonstrate that the quantum noise in this ``diluted'' noise model can already significantly impact the output of the quantum algorithms. 

\section{Additional numerical results}\label{app:add-num}
\begin{figure}[htbp]
    \centering
    \vspace{-10pt}
    \subfigure[\label{appfig:ts-qasm-large-phi}]{
        \includegraphics[width=14cm]{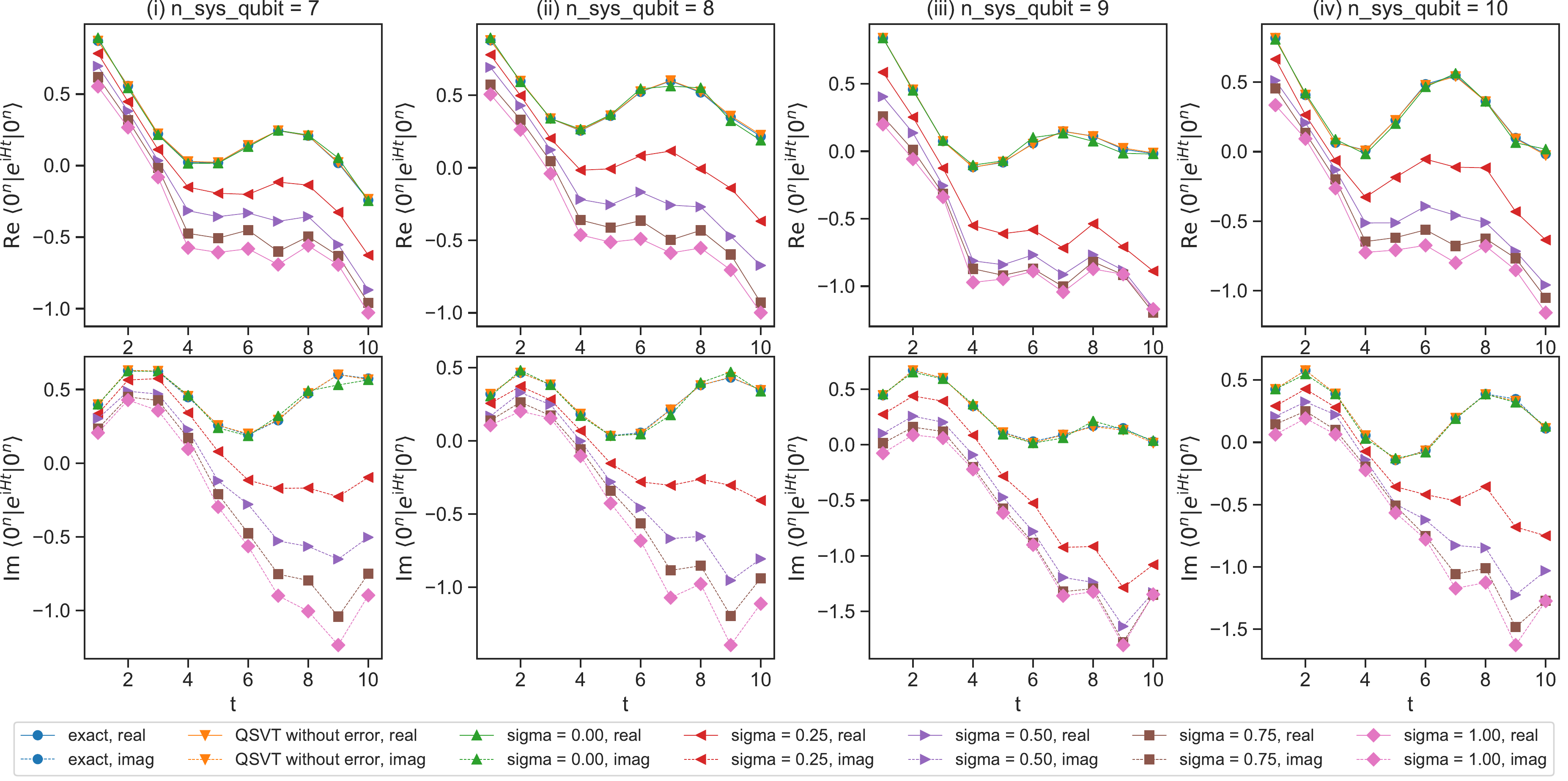}
    }
    
    \subfigure[\label{appfig:sm-qasm-large-phi}]{
        \includegraphics[width=14cm]{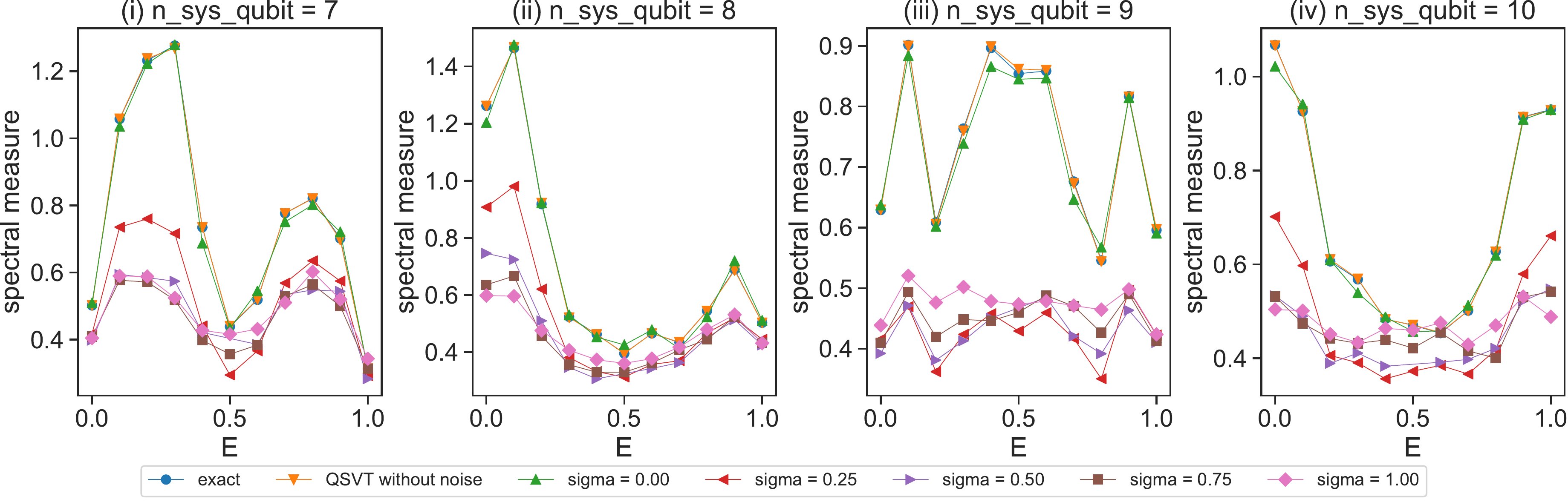}
    }
    
    \vspace{-10pt}
    \caption{Numerical results on QVM by using QSVT phase factors with larger length compared to those in the main text. Other parameters are set to be the same as those in the main text. (a) Time series. The details about phase factors are given in \cref{tab:param-ts-large}. (b) Spectral measure. The details about phase factors are given in \cref{tab:param-sm-large}.}
    \label{appfig:tradeoff-qsvt-length}
\end{figure}

As discussed in the main text, the circuit depth has a crucial impact on the accuracy of the output. For a given \bem, the circuit depth can be most effectively reduced by reducing the length of QSVT phase factors. Hence the choice of a proper length of QSVT phase factors needs to balance the error due to the polynomial approximation, and that caused by the noisy quantum device. To verify this, we increase the length of QSVT phase factors for computing the time series and spectral measures (details in \cref{tab:param-ts-large} and \cref{tab:param-sm-large}). From \cref{appfig:tradeoff-qsvt-length} it is evident that the numerical results are more accurate in the absence of noise when $\sigma=0$. However, as the noise magnitude increases, the results with a deeper quantum circuit become significantly worse compared to those in \cref{fig:ts} and \cref{fig:sm-qasm}.

\begin{figure}[htbp]
    \centering
    \subfigure[\label{appfig:ta-cma-ibmq}]{
        \includegraphics[width=14cm]{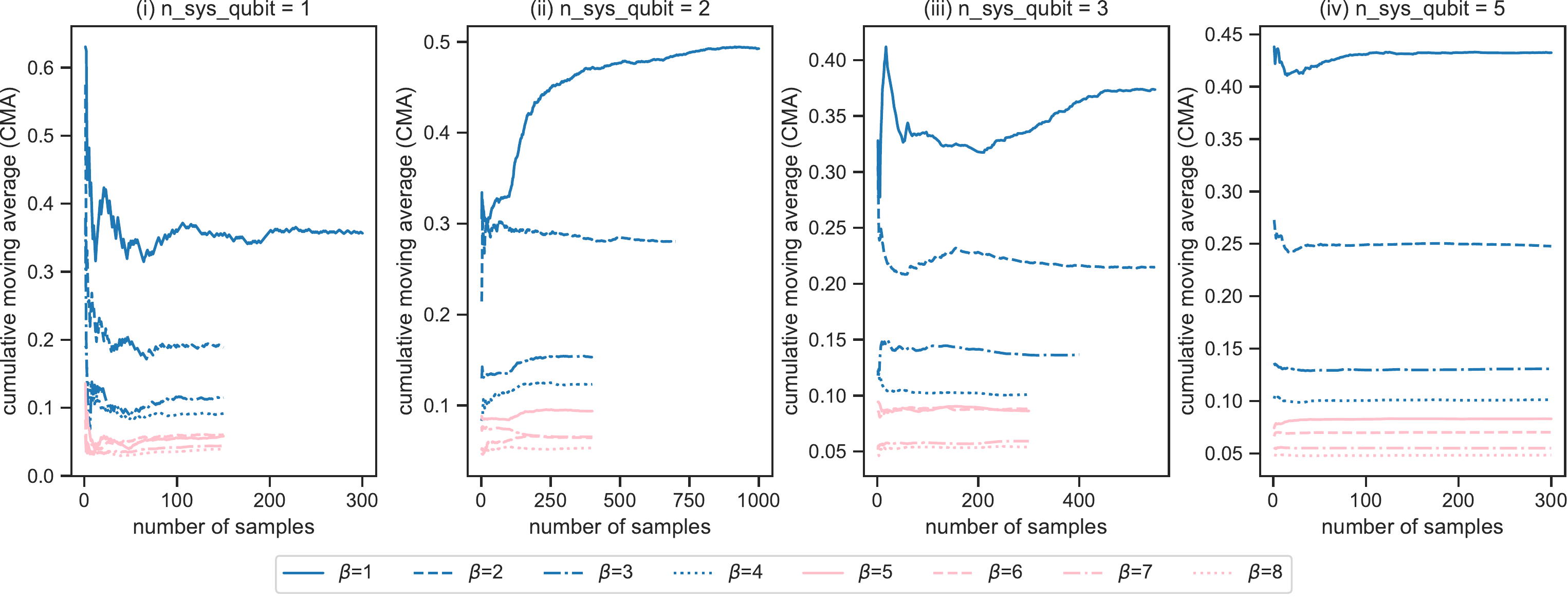}
    }
    
    \subfigure[\label{appfig:ta-cma-qasm}]{
        \includegraphics[width=14cm]{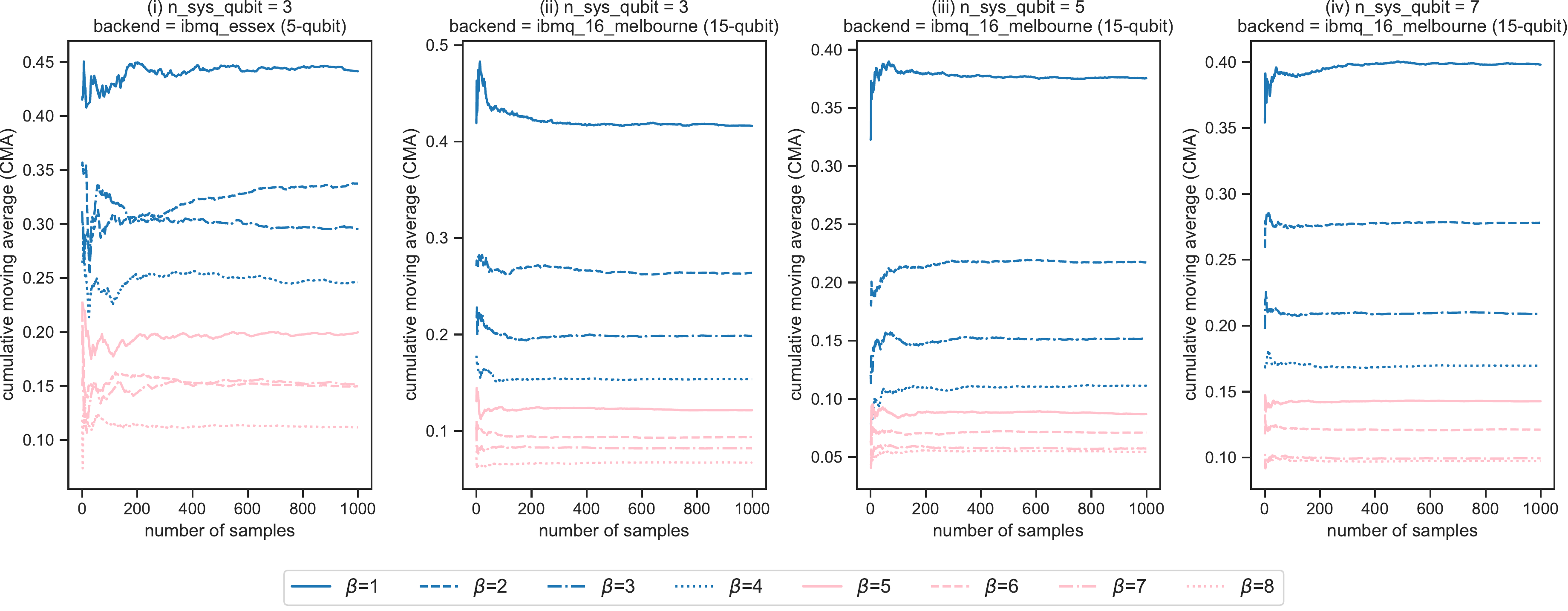}
    }
    \vspace{-10pt}
    \caption{Cumulative moving average (CMA) of the thermal average of the energy in the METTS algorithm. (a) CMA of the results computed on IBM Q backends. Enough samples are used for each parameter to guarantee the convergence of the result. (b) CMA of the results computed on QVM with noise magnitude $\sigma=1$.}
    \label{appfig:ta-cma}
\end{figure}

The METTS algorithm can be used to compute thermal averages via a Markov chain. The convergence behavior, plotted using the cumulative moving average (CMA) of the thermal average of the energy, is given in \cref{appfig:ta-cma}.

\section{Simulation details}\label{app:details}
\begin{figure}[htbp]
    \centering
    \includegraphics[width=0.4\textwidth]{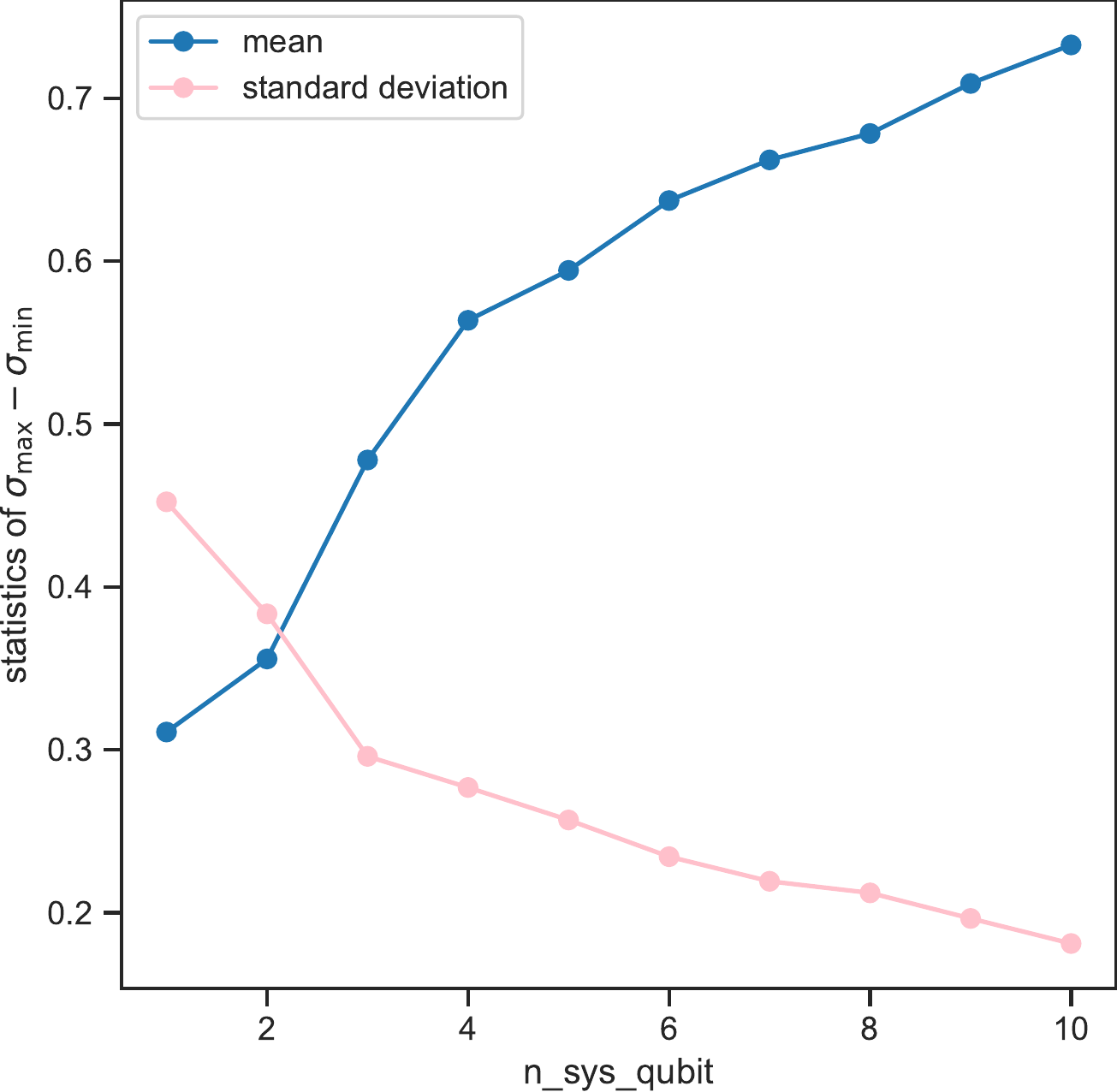}
    \caption{Singular-value distributions of different numbers of system qubits. The depth of the random circuit is adaptively chosen according to the number of system qubits. When the number of system qubits is less than or equal to $3$, the coupling map is retrieved from the 5-qubit backend \textsf{ibmq\_burlington}, otherwise, the coupling map is retrieved from the 15-qubit backend \textsf{ibmq\_16\_melbourne}. The probability of CNOT is 0.5 and the basic gate is restricted to $\Gamma = \{ \mathrm{U1}, \mathrm{U2}, \mathrm{CNOT} \}$. We generate 500 \bem's for each number of system qubits and compute the statistics of the difference between the maximal and minimal singular value of the block-encoded matrix.}
    \label{fig:svd-dist}
\end{figure}

\begin{figure}[htbp]
    \centering
    \subfigure[\label{fig:svd-ibmq}]{
        \includegraphics[width=0.7\textwidth]{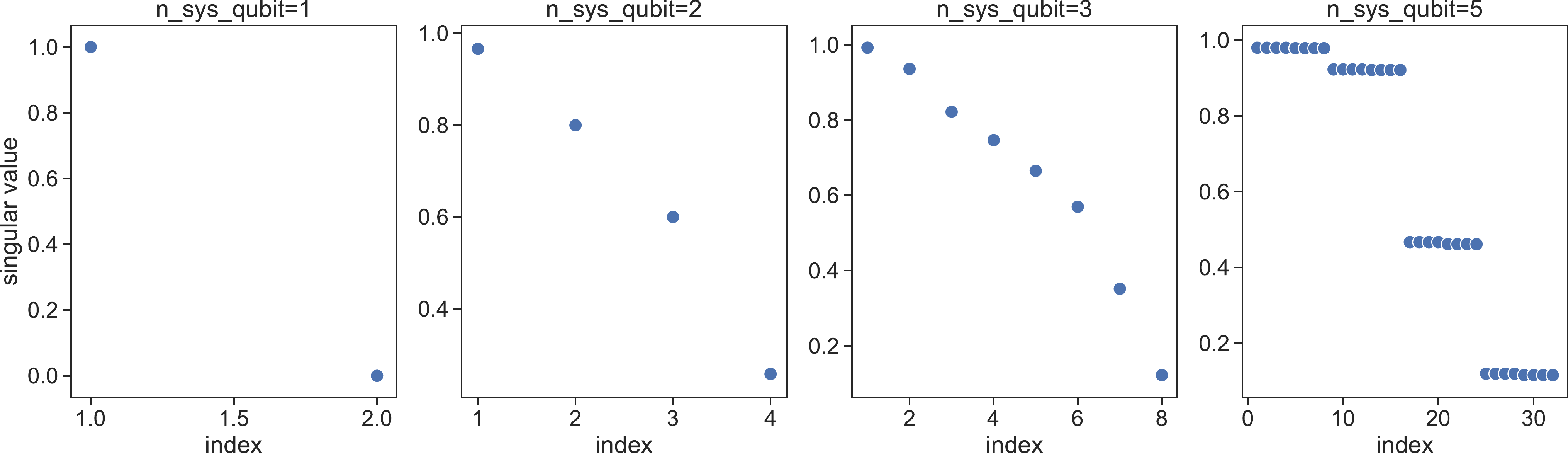}
    }

    \subfigure[\label{fig:svd-qvm}]{
        \includegraphics[width=0.7\textwidth]{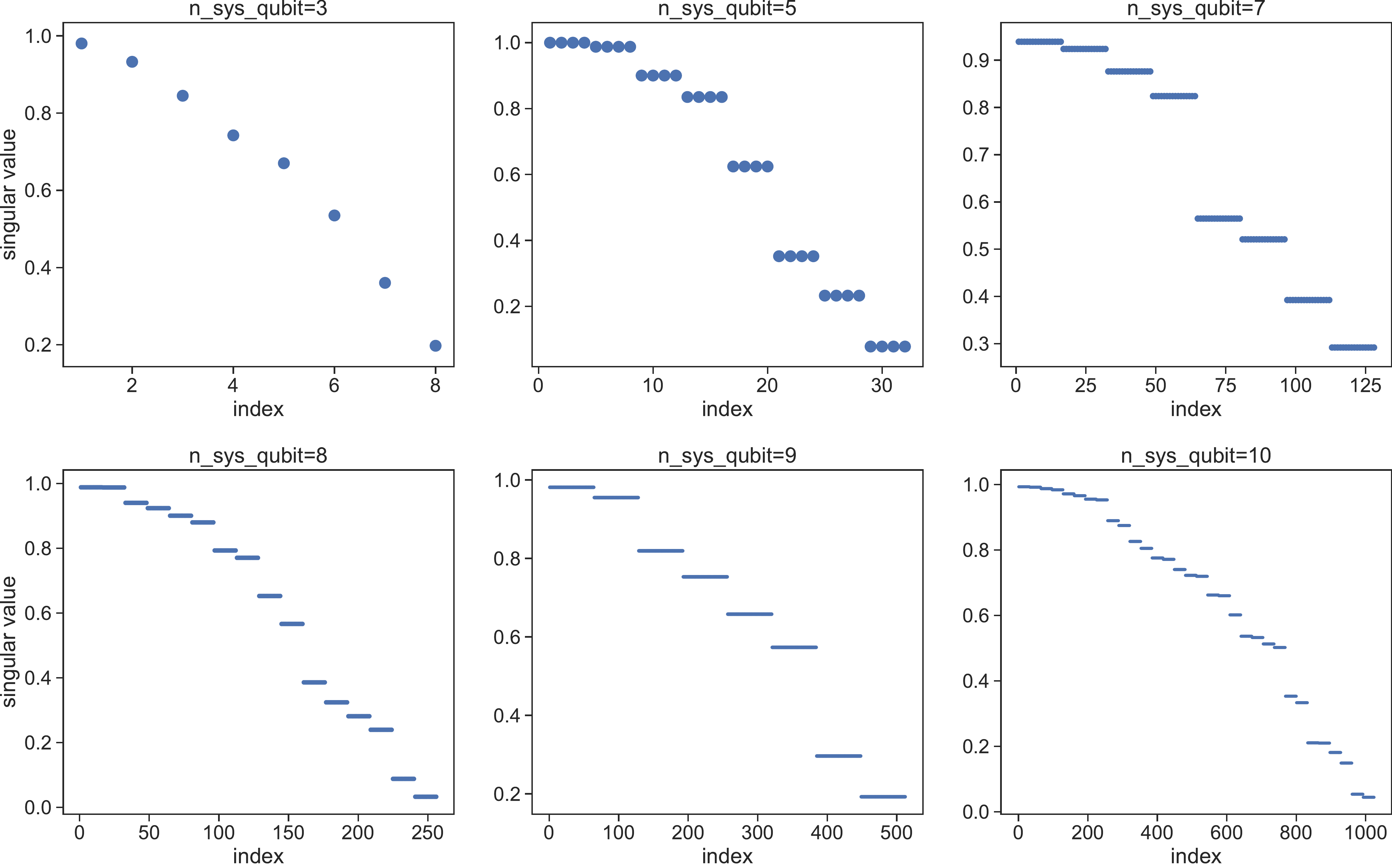}
    }
    \caption{Singular values of the matrices used in the numerical tests in the main text. The blue dots mark the singular values of the block-encoded matrix of a \bem circuit. When the number of system qubits is less than or equal to $3$, the coupling map is identified as that on the $5$-qubit backend \textsf{ibmq\_burlington}. Other results are obtained using the coupling map retrieved from the $15$-qubit backend \textsf{ibmq\_16\_melbourne}. (a) Singular values of matrices used in the numerical experiments run on quantum computing backends provided by IBM Q. The probability of CNOT is 0.1 when generating random circuits. (b) Singular values of matrices used in the numerical experiments run on QVM. The probability of CNOT is 0.5 when generating random circuits.}
    \label{fig:svd-matrix}
\end{figure}

In this section we provide the details of QSVT circuits used in our numerical experiments. The choice of the depth $\ell$ of the random quantum circuit for different number of system qubits $n$ is discussed in the main text. The logical gate count is given by the number of basic gates used in implementing the circuit, where we set the basic gate set to $\Gamma = \{ \mathrm{U1}, \mathrm{U2}, \mathrm{CNOT} \}$ in numerical experiments. We display the statistics of singular-value distributions of distinct numbers of system qubits in \cref{fig:svd-dist}. From the mean and standard deviation, it is evident that under our choice of the parameters of random circuit, the singular values of the block-encoded matrices varies largely. The singular values of relevant matrices which are used in the numerical tests in the main text are displayed in \cref{fig:svd-matrix}. Those matrices are block-encoded in \bem circuits generated according to our setup and \cref{alg:custom-rqc} at random. Such results indicate that the \bem model can generate at least non-trivial block-encoded matrices useful for testing the performance of quantum algorithms for numerical linear algebra tasks.

The details of the parameters used in the QSVT circuits are given below. We introduce a scaling factor, which is slightly greater than the maximal absolute value of the target polynomial, to scale the target polynomial so that it can be properly parameterized as QSVT phase factors.

\begin{table}[htbp]
    \centering
    \scalebox{0.75}{\begin{tabular}{|c|c|c|c|c|c|}
    \hline
     & $\kappa$ & length of QSVT phase factors $\qspdeg + 1$ & QSVT approximation error & logical gate count & scaling factor\\
    \hline
    \multirow{2}{*}{IBM Q} & 2 & 3 & 2.79722e-02 & $23 + 2 \cdot \ell \cdot n$ & 3.59306\\
     & 2 & 11 & 2.44481e-05 & $79 + 10 \cdot \ell \cdot n$ & 3.59306\\
     \hline
    \multirow{4}{*}{QVM} & 2 & 5 & 6.18245e-03 & $37 + 4 \cdot \ell \cdot n$ & 2.38234\\
     & 5 & 7 & 1.90152e-02 & $51 + 6 \cdot \ell \cdot n$ & 5.86631\\
     & 10 & 13 & 7.45462e-03 & $93 + 12 \cdot \ell \cdot n$ & 11.89390\\
     & 20 & 19 & 6.65999e-03 & $135 + 18 \cdot \ell \cdot n$ & 23.81003\\
    \hline
    \end{tabular}}
    \caption{Parameters of the QSVT circuit in solving QLSP, i.e., in \cref{fig:ls-ibmq-benchmark} and \cref{fig:ls-qasm-kappa}.}
    \label{tab:param-qlsp}
\end{table}

\begin{table}[htbp]
    \centering
    \scalebox{0.675}{\begin{tabular}{|c|c|c|c|c|c|c|}
    \hline
    & $t$ & $\eta$ & length of QSVT phase factors $\qspdeg + 1$ & QSVT approximation error & logical gate count & scaling factor\\
    \hline
    \multirow{10}{*}{real part} & 1 & 1.0 & 3 & 1.23670e-02 & $23 + 2 \cdot \ell \cdot n$ & 1.21807\\
    & 2 & 1.0 & 3 & 4.25711e-02 & $23 + 2 \cdot \ell \cdot n$ & 1.26458\\
    & 3 & 1.0 & 5 & 9.64293e-03 & $37 + 4 \cdot \ell \cdot n$ & 1.21400\\
    & 4 & 1.5 & 7 & 8.48770e-03 & $51 + 6 \cdot \ell \cdot n$ & 1.34011\\
    & 5 & 2.0 & 7 & 2.66925e-02 & $51 + 6 \cdot \ell \cdot n$ & 1.51827\\
    & 6 & 1.5 & 9 & 2.10169e-02 & $65 + 8 \cdot \ell \cdot n$ & 1.35177\\
    & 7 & 1.5 & 9 & 3.47455e-02 & $65 + 8 \cdot \ell \cdot n$ & 1.39998\\
    & 8 & 1.5 & 9 & 5.78363e-02 & $65 + 8 \cdot \ell \cdot n$ & 1.44148\\
    & 9 & 1.5 & 11 & 2.84139e-02 & $79 + 10 \cdot \ell \cdot n$ & 1.37467\\
    & 10 & 1.5 & 11 & 3.26549e-02 & $79 + 10 \cdot \ell \cdot n$ & 1.38139\\
    \hline
    \multirow{10}{*}{imaginary part} & 1 & 1.0 & 3 & 1.14646e-02 & $23 + 2 \cdot \ell \cdot n$ & 1.16750\\
    & 2 & 1.0 & 3 & 4.73088e-02 & $23 + 2 \cdot \ell \cdot n$ & 1.24295\\
    & 3 & 1.0 & 5 & 1.60676e-03 & $37 + 4 \cdot \ell \cdot n$ & 1.19769\\
    & 4 & 1.0 & 5 & 8.83397e-03 & $37 + 4 \cdot \ell \cdot n$ & 1.18742\\
    & 5 & 1.5 & 5 & 7.76900e-02 & $37 + 4 \cdot \ell \cdot n$ & 1.23512\\
    & 6 & 1.5 & 7 & 3.15931e-02 & $51 + 6 \cdot \ell \cdot n$ & 1.36811\\
    & 7 & 1.5 & 7 & 6.47625e-02 & $51 + 6 \cdot \ell \cdot n$ & 1.35109\\
    & 8 & 1.5 & 9 & 5.50680e-02 & $65 + 8 \cdot \ell \cdot n$ & 1.43342\\
    & 9 & 1.5 & 9 & 5.73218e-02 & $65 + 8 \cdot \ell \cdot n$ & 1.40154\\
    & 10 & 1.5 & 11 & 6.46945e-02 & $79 + 10 \cdot \ell \cdot n$ & 1.41058\\
    \hline
    \end{tabular}}
    \caption{Parameters of  QSVT circuits in the time series analysis, i.e., in \cref{fig:ts}.}
    \label{tab:param-ts}
\end{table}

\begin{table}[htbp]
    \centering
    \scalebox{0.675}{\begin{tabular}{|c|c|c|c|c|c|c|}
    \hline
    & $t$ & $\eta$ & length of QSVT phase factors $\qspdeg + 1$ & QSVT approximation error & logical gate count & scaling factor\\
    \hline
    \multirow{10}{*}{real part} & 1 & 1.0 & 5  & 1.35882e-04 & $37 + 4 \cdot \ell \cdot n$ & 1.20020\\
    & 2 & 1.0 & 5 & 2.07363e-03 & $37 + 4 \cdot \ell \cdot n$ & 1.20298\\
    & 3 & 1.0 & 7 & 9.83304e-04 & $51 + 6 \cdot \ell \cdot n$ & 1.19894\\
    & 4 & 2.0 & 9 & 4.56577e-03 & $65 + 8 \cdot \ell \cdot n$ & 1.46804\\
    & 5 & 2.0 & 11 & 3.37127e-03 & $79 + 10 \cdot \ell \cdot n$ & 1.46510\\
    & 6 & 2.0 & 13 & 3.17200e-03 & $ 93 + 12 \cdot \ell \cdot n$ & 1.47531\\
    & 7 & 2.5 & 13 & 4.32981e-03 & $93 + 12 \cdot \ell \cdot n$ & 1.58693\\
    & 8 & 2.0 & 15 & 4.40561e-03 & $107 + 14 \cdot \ell \cdot n$ & 1.47748\\
    & 9 & 2.0 & 17 & 5.14176e-03 & $121 + 16 \cdot \ell \cdot n$ & 1.47880\\
    & 10 & 2.5 & 17 & 5.05998e-03 & $121 + 16 \cdot \ell \cdot n$ & 1.59715\\
    \hline
    \multirow{10}{*}{imaginary part} & 1 & 1.0 & 5 & 2.88576e-04 & $37 + 4 \cdot \ell \cdot n$ & 1.15186\\
    & 2 & 1.0 & 5 & 1.25689e-03 & $37 + 4 \cdot \ell \cdot n$ & 1.19857\\
    & 3 & 1.0 & 5 & 1.60676e-03 & $37 + 4 \cdot \ell \cdot n$ & 1.19769\\
    & 4 & 1.0 & 7 & 4.05038e-03 & $51 + 6 \cdot \ell \cdot n$ & 1.19621\\
    & 5 & 1.5 & 9 & 3.23071e-03 & $65 + 8 \cdot \ell \cdot n$ & 1.34575\\
    & 6 & 2.0 & 11 & 6.19908e-03 & $79 + 10 \cdot \ell \cdot n$ & 1.48017\\
    & 7 & 3.0 & 13 & 3.50032e-03 & $93 + 12 \cdot \ell \cdot n$ & 1.70412\\
    & 8 & 3.0 & 15 & 2.35584e-03 & $107 + 14 \cdot \ell \cdot n$ & 1.70038\\
    & 9 & 4.0 & 15 & 3.59205e-03 & $107 + 14 \cdot \ell \cdot n$ & 1.90551\\
    & 10 & 3.0 & 17 & 3.53947e-03 & $121 + 16 \cdot \ell \cdot n$ & 1.70218\\
    \hline
    \end{tabular}}
    \caption{Parameters of the QSVT circuit in time series analysis with higher approximation precision, i.e., in \cref{appfig:ts-qasm-large-phi}.}
    \label{tab:param-ts-large}
\end{table}

\begin{table}[htbp]
    \centering
    \scalebox{0.8}{\begin{tabular}{|c|c|c|c|c|}
    \hline
    $E$ &length of QSVT phase factors $\qspdeg + 1$ & QSVT approximation error & logical gate count & scaling factor\\
    \hline
    0.0 & 11 & 3.20242e-02 & $79 + 10 \cdot \ell \cdot n$ & 2.14095\\
    0.1 & 11 & 5.59268e-02 & $79 + 10 \cdot \ell \cdot n$ & 2.14095\\
    0.2 & 11 & 1.23926e-01 & $79 + 10 \cdot \ell \cdot n$ & 2.14095\\
    0.3 & 11 & 1.32550e-01 & $79 + 10 \cdot \ell \cdot n$ & 2.14095\\
    0.4 & 11 & 1.36737e-01 & $79 + 10 \cdot \ell \cdot n$ & 2.14095\\
    0.5 & 11 & 1.71503e-01 & $79 + 10 \cdot \ell \cdot n$ & 2.14095\\
    0.6 & 11 & 1.36727e-01 & $79 + 10 \cdot \ell \cdot n$ & 2.14095\\
    0.7 & 11 & 1.32529e-01 & $79 + 10 \cdot \ell \cdot n$ & 2.14095\\
    0.8 & 11 & 1.23922e-01 & $79 + 10 \cdot \ell \cdot n$ & 2.14095\\
    0.9 & 11 & 5.59258e-02 & $79 + 10 \cdot \ell \cdot n$ & 2.14095\\
    1.0 & 11 & 3.20242e-02 & $79 + 10 \cdot \ell \cdot n$ & 2.14095\\
    \hline
    \end{tabular}}
    \caption{Parameters of the QSVT circuit in computing spectral measure, i.e., in \cref{fig:sm-qasm}.}
    \label{tab:param-sm}
\end{table}

\begin{table}[htbp]
    \centering
    \scalebox{0.8}{\begin{tabular}{|c|c|c|c|c|}
    \hline
    $E$ &length of QSVT phase factors $\qspdeg + 1$ & QSVT approximation error & logical gate count & scaling factor\\
    \hline
    0.0 & 21 & 9.59006e-04 & $149 + 20 \cdot \ell \cdot n$ & 2.14095\\
    0.1 & 27 & 5.02000e-03 & $191 + 26 \cdot \ell \cdot n$ & 2.14095\\
    0.2 & 33 & 5.48205e-03 & $233 + 32 \cdot \ell \cdot n$ & 2.14095\\
    0.3 & 37 & 5.90586e-03 & $261 + 36 \cdot \ell \cdot n$ & 2.14095\\
    0.4 & 41 & 4.66722e-03 & $289 + 40 \cdot \ell \cdot n$ & 2.14095\\
    0.5 & 43 & 4.38167e-03 & $303 + 42 \cdot \ell \cdot n$ & 2.14095\\
    0.6 & 41 & 4.66727e-03 & $289 + 40 \cdot \ell \cdot n$ & 2.14095\\
    0.7 & 39 & 3.70179e-03 & $275 + 38 \cdot \ell \cdot n$ & 2.14095\\
    0.8 & 35 & 3.29638e-03 & $247 + 34 \cdot \ell \cdot n$ & 2.14095\\
    0.9 & 27 & 5.01941e-03 & $191 + 26 \cdot \ell \cdot n$ & 2.14095\\
    1.0 & 19 & 2.53240e-03 & $135 + 18 \cdot \ell \cdot n$ & 2.14095\\
    \hline
    \end{tabular}}
    \caption{Parameters of the QSVT circuit in computing spectral measure with higher approximation precision, i.e., in \cref{appfig:sm-qasm-large-phi}.}
    \label{tab:param-sm-large}
\end{table}

\begin{table}[htbp]
    \centering
    \scalebox{0.725}{\begin{tabular}{|c|c|c|c|c|c|}
    \hline
    & $\beta$ & length of QSVT phase factors $\qspdeg + 1$ & QSVT approximation error & logical gate count & scaling factor\\
    \hline
    \multirow{8}{*}{numerator} & 1 & 4 & 8.10003e-03 & $30 + 3 \cdot \ell \cdot n$ & 0.72602\\
    & 2 & 4 & 3.35247e-02 & $30 + 3 \cdot \ell \cdot n$ & 0.53351\\
    & 3 & 6 & 9.27231e-03 & $44 + 5 \cdot \ell \cdot n$ & 0.42483\\
    & 4 & 6 & 1.96455e-02 & $44 + 5 \cdot \ell \cdot n$ & 0.37029\\
    & 5 & 6 & 3.36597e-02 & $44 + 5 \cdot \ell \cdot n$ & 0.33182\\
    & 6 & 8 & 9.57788e-03 & $58 + 7 \cdot \ell \cdot n$ & 0.29986\\
    & 7 & 8 & 1.51955e-02 & $58 + 7 \cdot \ell \cdot n$ & 0.27823\\
    & 8 & 8 & 2.21488e-02 & $58 + 7 \cdot \ell \cdot n$ & 0.26052\\
    \hline
    \multirow{8}{*}{denominator} & 1 & 3 & 1.03401e-02 & $23 + 2 \cdot \ell \cdot n$ & 1.18530\\
    & 2 & 3 & 3.37925e-02 & $23 + 2 \cdot \ell \cdot n$ & 1.15324\\
    & 3 & 5 & 7.30555e-03 & $37 + 4 \cdot \ell \cdot n$ & 1.18960\\
    & 4 & 5 & 1.40523e-02 & $37 + 4 \cdot \ell \cdot n$ & 1.18014\\
    & 5 & 5 & 2.25225e-02 & $37 + 4 \cdot \ell \cdot n$ & 1.16846\\
    & 6 & 5 & 3.22698e-02 & $37 + 4 \cdot \ell \cdot n$ & 1.15529\\
    & 7 & 7 & 8.56996e-03 & $51 + 6 \cdot \ell \cdot n$ & 1.18781\\
    & 8 & 7 & 1.20692e-02 & $51 + 6 \cdot \ell \cdot n$ & 1.18290\\
    \hline
    \end{tabular}}
    \caption{Parameters of the QSVT circuit in computing the thermal average of the energy, i.e., in \cref{fig:ta}.}
    \label{tab:param-ta}
\end{table}

\end{document}